\documentstyle[12pt,psfig,epsfig]{article}
\def\be{\begin{equation}}
\def\ee{\end{equation}}
\def\ba{\begin{eqnarray}}
\def\ea{\end{eqnarray}}
\def\ga{\mathrel{\raise.3ex\hbox{$>$\kern-.75em\lower1ex\hbox{$\sim$}}}}
\def\la{\mathrel{\raise.3ex\hbox{$<$\kern-.75em\lower1ex\hbox{$\sim$}}}}

\def\simgt{\mathrel{\raise.3ex\hbox{$>$\kern-.75em\lower1ex\hbox{$\sim$}}}}
\def\simlt{\mathrel{\raise.3ex\hbox{$<$\kern-.75em\lower1ex\hbox{$\sim$}}}}

\newcommand{\bi}[1]{\bibitem{#1}}
\newcommand{\fr}[2]{\frac{#1}{#2}}

\newcommand{\qq}{\langle \ov{q}q\rangle}
\newcommand{\uGu}{\bar{u}g_s(G\si) u}
\newcommand{\dGd}{\bar{d}g_s(G\si) d}
\newcommand{\nc}{\newcommand}
\newcommand{\uu}{\bar{u}u}
\newcommand{\dd}{\bar{d}d}
\nc{\gone}{\bar g_{\pi NN}^{(1)}}
\nc{\gzero}{\bar g_{\pi NN}^{(0)}}
\nc{\al}{\alpha}
\nc{\de}{\delta}
\nc{\ep}{\epsilon}
\nc{\ze}{\zeta}
\nc{\et}{\eta}
\renewcommand{\th}{\theta}
\nc{\Th}{\Theta}
\nc{\ka}{\kappa}
\nc{\rh}{\rho}
\nc{\si}{\sigma}
\nc{\ta}{\tau}
\nc{\up}{\upsilon}
\nc{\ph}{\phi}
\nc{\ch}{\chi}
\nc{\ps}{\psi}
\nc{\om}{\omega}
\nc{\Ga}{\Gamma}
\nc{\De}{\Delta}
\nc{\La}{\Lambda}
\nc{\Si}{\Sigma}
\nc{\Up}{\Upsilon}
\nc{\Ph}{\Phi}
\nc{\Ps}{\Psi}
\nc{\Om}{\Omega}
\nc{\ptl}{\partial}
\nc{\del}{\nabla}
\nc{\ov}{\overline}
\nc{\newcaption}[1]{\centerline{\parbox{15cm}{\caption{#1}}}}

\begin{document}

\begin{titlepage}
\rightline{UMN--TH--2222/03}
\rightline{FTPI--MINN--03/34}
\rightline{DAMTP--2003--126}
\rightline{UVIC--TH--03/09}
\rightline{DESY 03-187}
\rightline{hep-ph/0311314}
\rightline{November 2003}
\begin{center}
%\vspace{0.5cm}

\Large {\bf Electric Dipole Moments in the MSSM at Large 
\boldmath{$\tan\beta$}}\\[3mm]

\vspace*{0.5cm}
\normalsize

{\bf  Durmus Demir$^{\,(a)}$, Oleg Lebedev$^{\,(b)}$, 
Keith A. Olive$^{\,(a)}$}, {\bf Maxim Pospelov$^{\,(c,d)}$} and
{\bf Adam Ritz$^{\,(e)}$}

\smallskip
\medskip

$^{\,(a)}${\it Theoretical Physics Institute, School of Physics and
Astronomy,\\  University of Minnesota, Minneapolis, MN 55455, USA}

$^{\,(b)}${\it DESY Theory Group, D-22603 Hamburg, Germany}

$^{\,(c)}${\it Centre for Theoretical Physics,
                                 CPES,
                                 University of Sussex,
                                 Brighton BN1~9QJ, UK }

$^{\,(d)}${\it Deptartment of Physics and Astronomy, University of Victoria, \\
     Victoria, BC, V8P 1A1 Canada}

$^{\, (e)} ${\it Department of Applied Mathematics and Theoretical
         Physics, \\ Centre for Mathematical Sciences,\\ University of
         Cambridge, Wilberforce Rd., Cambridge CB3 0WA, UK}
\smallskip
\end{center}
\vskip0.2in

\centerline{\large\bf Abstract}

Within the minimal supersymmetric standard model (MSSM), the
large  $\tan\beta$  regime can lead to important modifications in the
pattern of CP-violating sources contributing to 
low energy electric dipole moments (EDMs). In particular, 
four-fermion CP-violating interactions induced by Higgs exchange 
should be accounted for alongside the constituent
EDMs of quarks and electrons. To this end, we present a comprehensive 
analysis of
three low energy EDM observables -- namely the EDMs of thallium, mercury 
and the 
neutron -- at large  $\tan\beta$, in terms of one- and two-loop 
contributions to 
the constituent EDMs and four-fermion interactions. We concentrate on 
the constrained MSSM as well as  the MSSM with non-universal Higgs masses,
and include the CP-violating phases of  $\mu$ and $A$.
Our results indicate that the atomic EDMs receive significant corrections
from four-fermion operators, especially when Im$(A)$ is the only
CP-violating source, whereas the neutron EDM remains relatively 
insensitive to these effects. As a consequence, in  a large portion of the
parameter space, one cannot infer a separate bound on the electron 
EDM via the experimental constraint on the thallium EDM. 
Furthermore,
we find that the electron EDM can be greatly reduced due to the
destructive interference of one- and two-loop contributions with the
latter being dominated by virtual staus.

\vspace*{2mm}
%\smallskip\newline

\end{titlepage}

\section{Introduction}

Electric dipole moments (EDMs) of the neutron \cite{n} and heavy 
atoms and molecules \cite{Tl,Hg,TlF,Xe,Cs,YbF,PbO} are primary  
observables in testing for flavor-neutral CP violation. 
The high degree of precision with which various experiments put 
limits on possible EDMs translates into 
stringent constraints on a variety of extensions of the Standard Model 
at and above the electroweak scale (see, e.g. \cite{KL}). The experiments 
that currently champion the best bounds on CP-violating parameters are the 
atomic EDMs of thallium and mercury and that of the neutron:
\ba
|d_{\rm Tl}| &<& 9 \times 10^{-25} e\, {\rm cm} \nonumber\\
|d_{\rm Hg}| &<& 2   \times 10^{-28}  e\, {\rm cm}     \\
|d_n|  &<&   6\times 10^{-26} e\, {\rm cm}.\nonumber
\label{explimit}
\ea

In this paper we address the theory of electric dipole moments in 
supersymmetric (SUSY) models with a large ratio of Higgs vacuum 
expectation values, or 
$\tan\beta$. Models with large $\tan\beta$ have recently received significant 
attention,  
stimulated in part by the final LEP results that  
impose significant constraints 
on the Higgs sector of the MSSM and imply a relatively large $\tan\beta$, 
$\tan\beta \ga 5$ \cite{LEPref}. Another motivation for large 
$\tan\beta$ stems from unified theories, which allow for the 
unification of top and bottom Yukawa couplings when 
$\tan\beta \sim 50$ \cite{Yunif}.

Supersymmetric models with the minimal field content (MSSM) allow for the 
presence of several CP-violating phases even in the most restrictive 
ansatz of flavor universality in the squark and slepton sectors. 
The null experimental EDM results pose a serious problem for the 
MSSM with superpartner masses around the electroweak 
scale. Indeed, a typical 
CP-violating SUSY phase of order one combined with 
${\cal O}(100{\rm~ GeV} - 1 {\rm ~ TeV})$ 
masses for superpartners would violate experimental constraints by up to 
three orders of magnitude \cite{WWI}. 
This generally requires the CP-violating phases to be very small,
unless there are cancellations among different contributions \cite{WWII,IN,WWIIa}
but these are largely disfavored by the mercury EDM constraint
\cite{FOPR,Barger:2001nu,Abel:2001mc}.
In string models, there are further complications since large EDMs are generally
induced even if the SUSY breaking dynamics conserve CP (in the sense that  the
SUSY breaking F--terms are all real) \cite{Abel:2001cv}.  Suppression of
these unwanted effects requires rather special circumstances, e.g.
dilaton domination with an approximate axial symmetry \cite{Lebedev:2002zt}.
Other options to avoid overproduction of EDMs in SUSY models include heavy sfermions
\cite{Nath:dn}, the presence of extra dimensions \cite{Kubo:2002pv},
additional symmetries \cite{Mohapatra:1997hi,susystuff}, etc. 
Yet, despite
the sizeable literature on EDMs in supersymmetric models, EDM constraints at 
large $\tan\beta$ remain poorly explored.

With this situation in mind, it is natural to ask what changes
to the EDM predictions will occur at large $\tan\beta$? One simple
observation is that the EDMs of the down quarks and the electrons, as 
induced by SUSY CP-violating phases, tend to grow because of their
dependence on the down-type Yukawa couplings which translates into a 
linear enhancement in $\tan\beta$. The second and in some sense 
more profound change was for a long time overlooked in the 
SUSY--EDM literature. At large $\tan\beta$, the observable EDMs of 
neutrons and heavy atoms receive contributions 
not only from the EDMs of the constituent particles, 
such as electrons and quarks,
but also from CP-odd {\it four-fermion operators}. The latter operators, 
since they are induced by Higgs exchange, receive an even more
significant enhancement by several powers of $\tan \beta$. The necessity of 
including these operators in the EDM analysis was shown 
by Barr \cite{Barr} for the case of a two Higgs 
doublet model with spontaneous breaking of CP, and recently by Lebedev and 
Pospelov \cite{LP} for the MSSM with explicit CP violation. 

In this paper, we present a thorough analysis of EDMs in the large 
$\tan\beta$ regime. We analyze the predictions for the three 
observables that champion the best constraints on the 
CP-odd sector of the theory: the atomic EDMs of thallium and 
mercury \cite{Tl,Hg}, and the neutron EDM \cite{n}. The restrictive 
assumption of Ref. \cite{LP} on the parametrically large 
superpartner mass scale will be lifted in 
this paper, and the effects of the four-fermion operators 
are analyzed alongside the usual electron and quark EDM and color EDM (CEDM) 
contributions. We first consider a framework in which
the scalar and gaugino masses, $m_0$ and $m_{1/2}$, as well as 
the tri-linear terms $A_0$, are unified at the GUT scale, 
i.e. the constrained MSSM (CMSSM) \cite{cmssm1,cmssm2}. 
In the CMSSM, the Higgs mixing mass, $\mu$, and the 
pseudoscalar Higgs mass, $m_A$, are determined by the
GUT scale parameters, $m_{1/2}, m_0, A_0$, and $\tan \beta$ using the radiative 
electroweak symmetry conditions.  The sign of $\mu$ is also 
free in the CP-conserving version of the CMSSM. 
We then repeat the 
same analysis for the case when scalar universality is relaxed in 
the Higgs sector (NUHM) \cite{nuhm1,nuhm2}. In this case, the 
Higgs soft masses $m_1$ and $m_2$ can be chosen independently of the
squark and slepton masses, which remain unified at the GUT scale. 
Alternatively, one may choose $\mu$ and $m_A$ as free parameters
and use the RGEs to determine $m_1$ and $m_2$ at the GUT scale. 
The positivity of $m_1^2$ and $m_2^2$ at the GUT scale may
impose a restriction on the parameter space in this case.

The universality of 
the gaugino masses at the GUT scale limits the 
number of physical SUSY CP-violating phases to 
two\footnote{This is a consequence of the U(1)${_R}$ and Peccei--Quinn
symmetries which allow us to eliminate two CP-violating phases (those of $B\mu$ and gaugino masses).
In the flavor--nonuniversal case, some of the phases can also be rotated away
by  redefining the quark superfields, such that the physical SUSY CP-violating phases
are reparametrization invariant combinations of the SUSY and SM flavor structures
\cite{Lebedev:2002wq}.}, which in the real 
gaugino mass basis can be identified with the 
phases of the $\mu$ and $A$ parameters. 
We perform the analysis separately for these two phases, as the scaling of 
the EDMs with $\tan\beta$ is quite different in each case.

The paper is organized as follows. In the next section we review the 
structure of the effective CP-odd Lagrangian at 1 GeV. We argue that  
at large $\tan \beta$ it should be modified by the 
presence of four-fermion operators. 
We then estimate the SUSY parameter regimes where 
these contributions become important sources for EDMs. Section 3 calculates the 
supersymmetric threshold corrections that induce CP violation via Higgs 
exchange. Section 4 addresses the modifications  of the 
thallium, mercury and neutron EDMs due to the presence of four-fermion 
operators and provides general formulae for these observables. In section 5 
we perform a numerical analysis of EDMs at $\tan \beta = 50$ in the CMSSM and
the NUHM. 
Section 6 concludes the paper by summarizing the modifications to the EDM 
bounds at large $\tan\beta$ and discussing  the  importance of different 
contributions to the EDMs. Some additional details on relevant hadronic 
calculations are given in appendices A and B.

\section{The structure of the low energy Lagrangian} 

We begin by recalling the standard structure of the effective CP-odd 
Lagrangian normalized at 1 GeV, which is the lowest scale allowing for
a perturbative quark-gluon description. 
It includes the theta term, the Weinberg operator, the EDMs of quarks and 
electrons, and the color EDMs of quarks: 
\begin{eqnarray}
\label{leff}
{\cal{L}}_{eff} &=& \frac{g_s^2}{32\pi^{2}}\ \bar\theta\
G^{a}_{\mu\nu} \widetilde{G}^{\mu\nu , a} +
\frac{1}{3} w\  f^{a b c} G^{a}_{\mu\nu} \widetilde{G}^{\nu \beta , b}
G^{~~ \mu , c}_{\beta} \nonumber\\  && -
\frac{i}{2} \sum_{i=e,u,d,s} d_i\ \overline{\psi}_i (F\sigma)\gamma_5 \psi  -
\frac{i}{2} \sum_{i=u,d,s} 
\widetilde{d}_i\ \overline{\psi}_i g_s (G\sigma)\gamma_5\psi + \cdots
\end{eqnarray}
For a given pattern of 
supersymmetry breaking $d_i$, $\tilde d_i$, and $w$ can be 
calculated explicitly, via one or two-loop SUSY diagrams, and evolved down 
to 1 GeV using perturbative QCD. On the other hand, only the additive radiative 
corrections to the theta term can be calculated, leaving $\bar\theta$ as a 
free parameter subject to an arbitrary initial condition. 
In supersymmetric models with CP-violating phases larger than 
$\sim 10^{-8}$ in the flavor-diagonal 
sector the only reasonable strategy to avoid the strong 
CP problem is to postulate the existence of the Peccei-Quinn 
relaxation mechanism \cite{PQ}, which removes the theta term from 
the list of contributing operators. 
We will therefore adopt this strategy and discard $\bar\theta$. 
When $\tan \beta$ is low, possible four fermion CP-odd operators 
(e.g. $(\bar qq)( \bar qi\gamma_5 q )$ and alike) can be safely neglected. 
Indeed, these operators involve a double flip of helicity associated with 
an additional $(m_q/v_{EW})^2$ suppression, and this makes these operators 
effectively of dimension 8, and thus totally negligible. 

The dependence of the experimentally measured EDMs of neutrons, 
and paramagnetic and diamagnetic atoms, on the Wilson coefficients 
in (\ref{leff}), as calculated at low 
$\tan\beta$, can be symbolically presented in the following way:
\begin{eqnarray}
\nonumber
d_{\rm Tl} &=&  d_{\rm Tl}(d_e)\\
\label{lowtanbeta}
 d_{n} &=&  d_{n}( \bar \theta , d_i, \tilde d_i, w)~~~~~~~~~~ {\rm at~low~}  \tan\beta .\\
d_{\rm Hg} &=&  d_{\rm Hg}(\bar \theta , \tilde d_i, d_i, w)\nonumber
\end{eqnarray}

The dependence of the EDMs of paramagnetic atoms and molecules 
(such as thallium) on the electron EDM alone makes them  the 
most valuable observables for SUSY models, as they allow us to 
link the experimental results directly to a 
calculable combination of CP-violating phases, and slepton and gaugino masses, 
while the uncertainty of atomic calculations enters as an overall factor. 
In fact, due to this property of paramagnetic EDMs, it is customary to quote 
the limit on $d_e$ directly, skipping the actual experimental limit on the 
EDM of the paramagnetic atom. The
EDMs of the neutron and mercury are far more complicated as a 
variety of hadronic (and in the case of mercury also nuclear and 
atomic)  matrix elements are needed in order translate the experimental 
results into limits on a specific combination of the Wilson coefficients.

The structure of the effective Lagrangian becomes even more complicated if 
$\tan \beta$ is taken to be a large parameter. Indeed, the operators 
induced by Higgs exchange that involve down-type quarks and 
charged leptons grow rapidly with $\tan\beta$ \cite{LP}. For now, we parametrize these 
operators by coefficients $C_{ij}$,
\begin{eqnarray}
 {\cal{L}}_{eff}^{4f} = \sum_{i,j = e,d,s,b} C_{ij} ~ (\bar \psi_i \psi_i)
(\bar \psi_j i \gamma_5 \psi_j),
 \label{4f}
\end{eqnarray} 
where $i,j$ run over flavor indices and the second index always indicates the 
fermion flavor that enters (\ref{4f}) via a pseudoscalar bilinear. 
These operators represent a subclass of the larger set of all 
possible CP-odd flavor-conserving 
four-fermion operators  considered in \cite{KKY,CPfourferm,PH}.
Note that the $b$-quark is heavy compared to the scale of 1 GeV and thus 
can be integrated out, producing new dimension 7 operators 
$\bar \psi_i \psi_i (G^a_{\mu\nu} \tilde G^{\mu\nu , a})$ and 
$\bar \psi_i i \gamma_5\psi_i (G^a_{\mu\nu} G^{\mu\nu , a})$ and 
leading to new 
higher loop contributions to $d_i$, $\tilde d_i$ and $w$. 
With this procedure in mind, we prefer to  keep the coefficients 
$C_{bi}$ and $C_{ib}$ explicitly so that integrating out the $b$-quarks is interpreted
as part of taking the corresponding matrix element. 
   
When the contributions from $C_{ij}$ are not negligible,  
the relation between the observables and the SUSY parameters encoded in 
the Wilson coefficients of Eqs.~(\ref{leff}) and (\ref{4f}) 
becomes considerably more involved:
\begin{eqnarray}
\nonumber
d_{\rm Tl} &= & d_{\rm Tl}(d_e, C_{de}, C_{se}, C_{be} )\\
\label{largetanbeta}
 d_{n} &=& d_{n}(\bar\theta, d_i, \tilde d_i, w, C_{q_1q_2})~~~ ~~~
{\rm at~large~}\tan\beta .\\
d_{\rm Hg}& = & d_{\rm Hg}(\bar\theta, d_i,\tilde d_i, w, C_{ij}).\nonumber
\end{eqnarray}
Paramagnetic EDMs lose their one-to-one relation to $d_e$, as 
they also become dependent on operators that involve quarks and 
gluons, and as a consequence are subject to considerable hadronic 
uncertainties.
Hence, at large $\tan\beta$, a separate bound on $d_e$ simply may not exist, 
and the bound on the EDM of the  paramagnetic atom 
(molecule) as a whole should be used instead. The 
mercury EDM receives a number of additional contributions 
from semileptonic and pure hadronic four-fermion operators. As a consequence, 
$(\tilde d_u - \tilde d_d)$ may no longer be the dominant contribution 
to $d_{\rm Hg}$ in contrast to the situation at low $\tan \beta$ \cite{FOPR}. 
Thus, the large $\tan\beta$ regime is qualitatively different from 
$\tan \beta \sim {\cal O}(1)$ and requires a separate 
dedicated analysis. 

To conclude this section, we will determine a combination of 
SUSY parameters that  regulates which 
effect is more important: one-loop EDMs or four-fermion operators. 
The required interpolating parameter can be deduced from
the asymptotics of the one-loop EDMs and four-fermion operators in the 
limit of a generic heavy superpartner mass scale $M$. 
For simplicity, and in this section only, 
we will not differentiate among different superpartners 
and assume the same mass for squarks, sleptons, gauginos, etc.  

Let us first consider a non-zero $\mu$--phase (relative to 
the gaugino mass). 
While the one-loop EDMs generally scale as 
$d_i/m_i \sim  \mu m_{\lambda}\tan\beta \sin \theta_\mu /M^4$,
the leading order contributions to $C_{ij}$ scale as 
$C \sim \mu m_{\lambda}(\tan\beta)^3 \sin \theta_\mu/(M^2 m_A^2)$ \cite{LP}, 
where $\theta_\mu$ is the CP-violating phase, $m_\lambda$ is a gaugino mass, 
and $M= {\rm max}(m_0,~ m_{1/2})$ stands for a sfermion or 
gaugino mass, whichever is larger.  Thus, we can introduce the
following combination of SUSY parameters that interpolates between 
the two regimes:
\be
\xi \equiv \frac{M}{m_A} \tan\beta .  
 \label{xi}
\ee
In the limit of large $M$, with $m_A$ fixed at the electroweak scale, 
all EDM-like observables are dominated by the 
four-fermion operators \cite{LP}. 
When $M$ is kept finite, there is a certain critical value for $\xi$ 
below which the usual one-loop EDMs dominate the observables, and above which 
the four-fermion operators are more important. 

Using the results of Refs. \cite{LP,FOPR}, we 
can estimate the critical value for 
$\xi$ in the limit when all superpartner masses are approximated by some  
heavy scale $m_0\sim m_{1/2}\sim M \ga v_{EW} \sim 246$ GeV, 
while $m_A$ is kept fixed near 
the electroweak scale. If all CP violation is induced by $\th_\mu$, 
the relative importance of the two sources is determined by the ratio  
\be
\frac{d_{\rm Tl}(C_{ij})}{ d_{\rm Tl}(d_e)} \simeq 10^{-5}\left[\frac{M}{m_A}
\tan\beta \right]^2 = 10^{-5} \xi^2, 
\label{comparison}
\ee
which suggests a critical value for $\xi$,
\be
\xi_{c} \sim 300  ~~~~~~ {\rm for } ~~~~~ \theta_\mu \neq 0.
\label{estim1}
\ee
From this estimate it becomes clear that when $m_A \simeq M$ the effects 
of the 
four-fermion operators are subleading but not negligible, reaching 20\%
of the $d_e$ contribution at maximal $\tan\beta$. On the other hand, even a 
mild hierarchy between the mass scales, $M \ga 5 m_A$, may lead 
to a contribution to atomic EDMs comparable to $d_e$, or even to 
dominance of the four-fermion contributions. 

In the case of the other CP-violating phase $\theta_A$,
the relative phase of the $A$ parameter and the gaugino mass, we estimate that 
the transition to dominance of the four-fermion operators takes place at 
\be
 \xi_c  \tan^{1/2} \beta  \sim 500~~~~~~ 
 {\rm for } ~~~~~ \theta_A \neq 0.
\label{estim2}
\ee
Here, the additional power of $\tan\beta$ arises from the fact that 
the one-loop induced $d_e(\theta_A)$ does not grow with $\tan \beta$. 
Thus in this case, even for $m_A\sim M$, the contribution 
of the four-fermion operators may become comparable to $d_e$. 

Both conditions (\ref{estim1}) and (\ref{estim2}) can be realized 
in a generic SUSY spectrum, but how realistic are they for the constrained 
MSSM and the MSSM with non-universal Higgs masses? 
We note that the simple comparison of 
$d_e$ and $C_{ij}$ made in this section may be  subject 
to significant modifications, 
as the masses of sleptons and squarks as well 
as other superpartners are affected  by the renormalization 
group evolution from the GUT scale to the electroweak scale. Therefore, we now turn 
to a numerical comparison of all contributions to the EDMs 
for different CP-violating phases and different patterns of SUSY breaking 
at large $\tan \beta$, as detailed in the next four sections.

\section{CP-odd four-fermion operators in the MSSM}

We now turn to  specific SUSY mechanisms that generate $C_{ij}$.
As shown in Ref.~\cite{LP}, $C_{ij}$ are generated primarily 
through the exchange of the pseudoscalar and 
scalar Higgs particles $A$ and $H$ with CP violation entering 
through vertex corrections. This dominates the one-loop induced  $C_{ij}$
due to large ${\cal O}(\tan^3 \beta)$ enhancements.  
Other contributions, such as CP-odd higher-loop effects, SUSY box diagrams, 
etc., can  have an even stronger dependence on $\tan \beta$. 
However, these effects are further suppressed either by 
additional loop factors  or by additional powers of small Yukawa couplings
and in this sense are subleading. Nevertheless, $A-H$ mixing may  
be numerically important \cite{LP}, and therefore its contribution has 
to be included in the EDM analysis.

\subsection{Higgs-mediated CP-odd four-fermion operators}

The $\tan\beta $ enhancement of 
certain loop corrections  in the 
down-type quark and charged lepton sectors
 is a well studied phenomenon \cite{vertexcorr}. 
Supersymmetric threshold corrections, such as those in Fig.~1, generate 
a new Yukawa interaction which is absent in the limit of 
exact supersymmetry. As a result, below the superpartner scale, 
the Yukawa interactions in the $D$-quark and 
charged lepton sectors are given by 
\ba
-{\cal L}_{Y} &=& (Y_{D}^{(0)}+Y_{D}^{(1)})
H_{1}D_{R}^\dagger D_{L}+{\cal Y}_{D}~H_{2}^{\dagger }
D^\dagger_{R}D_{L}\nonumber\\
 && +(Y_{E}^{(0)}
 +Y_{E}^{(1)})H_{1}E_{R}^\dagger E_{L}+{\cal Y}_{E}~H_{2}^{\dagger }
E_{R}^\dagger E_{L} +{\rm h.c.}\;,  \label{mass}
\ea
where $Y_{D,E}^{(0)}$ are the tree level Yukawa couplings  and
${\cal Y}_{D,E}$ and $Y_{D,E}^{(1)}$ are the one-loop-induced couplings. 
Thus, instead of being the Yukawa sector of a type II
two-Higgs doublet model, Eq.~(\ref{mass}) represents a generic two 
Higgs doublet model where Higgs exchange may lead to  violation of 
CP. 
One should keep in mind that Eq. (\ref{mass}) is valid in the limit of 
heavy superpartners. If the scale of SUSY breaking is close 
to the electroweak scale one should also include 
additional radiatively generated operators, e.g.
$(H_2^\dagger H_2)^nH_{2}^{\dagger }D^\dagger_{R}D_{L}$.

Due to flavor universality in the soft-breaking sector, 
flavor-changing effects can be ignored in EDM calculations, 
and a simple linear relation between $Y_{D,E}^{(0)}$ and  
${\cal Y}_{D,E}$ holds for every $D$ and $E$ flavor: 
\begin{equation}
{\cal Y}_{D}=J_{D}Y_{D}^{(0)},\;\;\;\;\;\;{\cal Y}_{E}=J_{E}Y_{E}^{(0)}\;.
\end{equation}
The calculation of the loop functions $J_{D}$ and $J_{E}$ 
is straightforward but tedious, as a large number of SUSY diagrams contribute.
Nevertheless, an analytical result for this set of loop diagrams is possible, 
and we present it in the next subsection.  

%%%%%%%%%%%%%%%%%%%%%%%
\begin{figure}
 \centerline{%
   \psfig{file=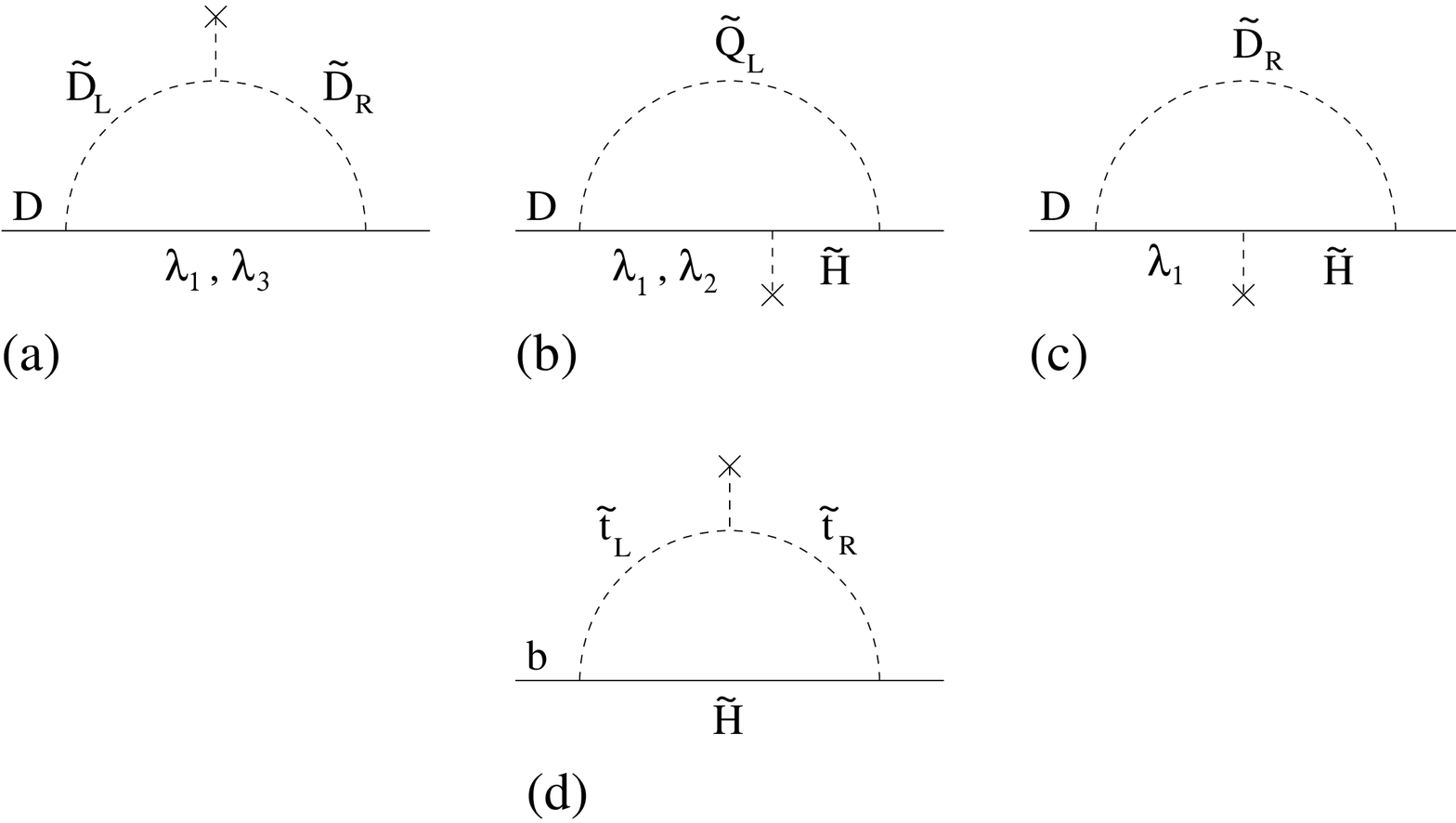,width=11.125cm,angle=0}%
         }
\vspace{0.1in}
 \newcaption{\small One-loop SUSY threshold corrections 
in the down quark and 
charged lepton Yukawa sectors. 
All four types of diagrams are relevant for the $b$-quark, 
while only (a)-(c) contribute 
in the case of $d$ and $s$ quarks, and the electron.
Here $\lambda_i$ denote the gauginos.
The vacuum insertion of $\langle H_2 \rangle $ indicated by a cross is for 
illustrative purposes only as all calculations are performed 
in the mass eigenstate basis. }
\end{figure}
%%%%%%%%%%%%%%%%%%%%%%%

SUSY CP violation arises from a mismatch between the phases of the $\mu$
and 
$A$ parameters, and $m_{1/2}$, and leads to complex masses and 
Yukawa couplings. 
The quark and lepton masses, 
\ba\label{mass1}
-{\cal L}_{\rm mass} &=& (Y_{D}^{(0)}+Y_{D}^{(1)})
\langle H_{1}\rangle D_{R}^\dagger D_{L}+{\cal Y}_{D}\langle H_{2}\rangle
D^\dagger_{R}D_{L}\nonumber\\
 && +(Y_{E}^{(0)}
 +Y_{E}^{(1)})\langle H_{1}\rangle E_{R}^\dagger E_{L}+
{\cal Y}_{E}\langle H_{2}\rangle
E_{R}^\dagger E_{L} +{\rm h.c.}\;,
\ea
can be made real via a chiral rotation of the $D$ and $E$ fields, 
\be
D_{R}\rightarrow e^{-i\delta ^{(m)}_{D}}D_{R},~~~~~ 
 E_{R}\rightarrow e^{-i\delta^{(m)}_{E}}E_{R},
\label{chrot}
\ee
where the rotation angles are defined by $\tan\beta$ and 
the loop functions $J_i$:
\be
\delta ^{(m)}_i=  
{\rm Arg}(Y_{i}^{(0)}+Y_{i}^{(1)}+{\cal Y}_i\tan \beta )\simeq         
{\rm Arg} (Y_i^{(0)}+ Y_i^{(1)}) +{\rm Arg}(1+J_i\tan \beta ),
\label{delta}
\ee
where $i=d,s,b,e$ are the flavors of interest.  $Y_{i}^{(1)}$ 
contains a loop  suppression factor and thus can be 
neglected. However, a large value of $\tan \beta $ can compensate 
the loop suppression of ${\cal Y}_i$ so that
the phases ${\rm Arg}(1+J_i\tan \beta )$ can be of order one. 
The vertices of the $A$ and $H$ Higgs bosons and  fermions, as derived 
from the Lagrangian (\ref{mass}), also contain a complex phase:
\be
\delta^{(H)}_i = {\rm Arg}( (Y_i^{(0)}+Y_i^{(1)})\tan\beta + {\cal Y}_i)\simeq
{\rm Arg} (Y_i^{(0)}+Y_i^{(1)}).
\ee
This equation is written under the assumption that 
$\tan\beta $ is a large parameter. 
The difference between these two phases,
\be
\delta_i = \delta ^{(m)}_i - \delta^{(H)}_i,
\label{phdiff}
\ee
constitutes a physical CP-violating phase that enters into the interactions 
of $H$ and $A$ with physical fermions $\psi_i$,
\be 
{\cal L}_{\rm CPV} = -\sum_{i=d,s,b,e} \fr{Y_i}{\sqrt{2}}\sin \delta_i 
 \left( \bar \psi_i \psi_i A + \bar \psi_i i\gamma_5 \psi_i H\right).
\ee
Note that the phase ${\rm Arg} (Y_i^{(0)}+Y_i^{(1)})$ drops out of the 
expression for $\delta_i$ as it should.
An exchange by the physical Higgses, as in Fig.~2,
will then produce CP--odd four--fermion interactions with the 
following coefficients 
%In case of 
%$C_{ij}$, the relevant interactions are induced by the exchange of a CP--odd Higgs boson $A$ 
%with CP violation entering through the quark-pseudoscalar vertex and by the exchange 
%and CP--even Higgs bosons $H$ and $h$ with CP violation on the electron line: 
\ba
{C}_{ij}^{\rm (vc)}(i,j=e,d,s,b) ~=~ -\frac{\tan ^{2}\beta }{2m_{A}^{2}} \, \frac{
Y_{i}^{SM}Y_{j}^{SM}\sin (\delta _{i}- \delta _{j})}{|1+J_{i}\tan \beta
||1+J_{j}\tan \beta |}.
\label{Cqe}
\ea
Here  $Y_{f}^{SM}$ denote the Standard Model values for the
Yukawa couplings, $Y_{f}^{SM}\equiv \sqrt{2}m_{f}/v$ ($v$=246 GeV), and the 
superscript of ${C}_{ij}^{\rm (vc)}$ signifies that this contribution originates
from a vertex correction. 
Note also that at leading order in $\tan\beta$ an exchange by
 the (lightest) $h$-Higgs 
does not contribute to $C_{ij}$, the 
difference between $m_{A}$ and $ m_{H}$ can be 
ignored, and the neutral Higgs mixing angle is trivial,
$\cos^{2}\alpha \simeq 1$. 

Examining expression (\ref{Cqe}), we notice immediately that 
\be
{C}_{ij}^{\rm (vc)}= - {C}_{ji}^{\rm (vc)},
\ee
and thus vertex corrections do not generate four-fermion operators 
that involve a single flavor such as, for example, 
$\bar dd \bar d i\gamma_5 d$. 

Combining  
Eqs.~(\ref{delta}) and (\ref{Cqe}), we observe that the coefficients 
$C_{ij}$ grow as  ${\cal O}(\tan^{3}\beta)$ as long 
as $J_{i(j)} \tan \beta < 1$, which is true for almost the entire available 
domain of $\tan\beta$. Moreover, the cubic growth of 
$C_{ij}$ generated at one loop due to  
explicit CP violation in the soft breaking sector is distinct from
the ${\cal O}(\tan ^{2}\beta)$-behavior of 
$C_{ij}$ in two-Higgs doublet models with 
spontaneous CP violation \cite{Barr}. 

%%%%%%%%%%%%%%%%%%%%%%%
\begin{figure}
 \centerline{%
   \psfig{file=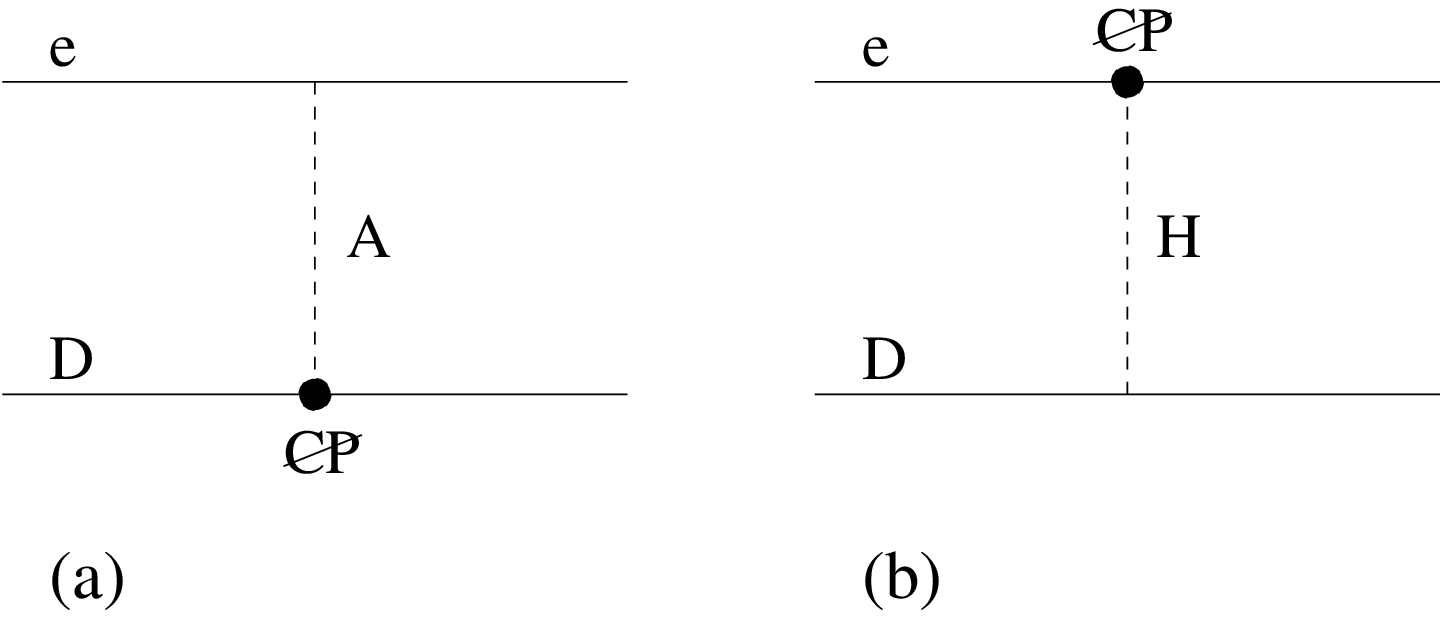,width=10cm,angle=0}%
         }
\vspace{0.1in}
 \newcaption{\small The 
Higgs-mediated electron-quark interaction $C_{qe}$ with  CP violation 
at the Higgs-quark vertex (a) and at the Higgs-electron vertex (b).}
\end{figure}
%%%%%%%%%%%%%%%%%%%%%%%

\subsubsection{Beyond the limit of heavy superpartners}

In order to go beyond the approximation of heavy superpartners, one should 
reformulate Eq. (\ref{Cqe}) in terms of the tree-level fermion 
masses $m_i^{(0)}= Y_i^{(0)}v_1/\sqrt{2}$ and the one-loop SUSY 
corrections to the fermion masses $\Delta m_i$, 
where {\em all operators} neglected in Eq. (\ref{mass}) are 
included\footnote{
A more complete approach would necessitate the use of the Coleman--Weinberg
effective potential. For our purposes, however, it suffices
to truncate the series of  non--renormalizable 
operators. }.
The one-loop-corrected mass terms can be made real by 
a phase redefinition (\ref{chrot}) with the phase now given by 
\begin{equation}
 \delta^{(m)}_i={\rm Arg}(1+ \Delta m_i/m_i^{(0)}) \;.
\label{mfull}
\end{equation}
Here we assume for convenience that $m_i^{(0)}$ is real. 
In the large $\tan\beta$ regime, the couplings of $A$ and $H$ 
Higgses to fermions can be obtained as a derivative 
of $m_i^{(0)}+\Delta m_i$ with respect to  $\langle H_1 \rangle = v_1$
at $v_1 =0$, 
and thus the phase  $\delta^{(H)}$ is now given by
\be
\delta^{\rm (H)}_i={\rm Arg} \left (\fr{\partial(m_i^{(0)} 
+ \Delta m_i)}{\partial v_1}\right).
\label{Hfull}
\ee
In the large $\tan\beta$ regime it is natural to expect that 
$\delta^{(H)}_i\ll \delta^{(m)}_i$, so that the physical CP-violating phase 
$\delta_i$ is given by ${\rm Arg}(1+ \Delta m_i/m_i^{(0)})$. 
However, in practice it often happens that $\delta^{(H)}_i$ 
cannot be neglected (see Sec.~5.1). 
Finally, the coefficients $1+J_i\tan\beta$ appearing in the denominator of 
Eq.~(\ref{Cqe}) should also be replaced 
by the expression $1+\Delta m_i/m_i^{(0)}$ 
(which is independent of $m_i^{(0)}$ because $\Delta m_i$ is proportional to 
$m_i^{(0)}$). With these modifications, Eq.(\ref{Cqe}) takes the form
\be
\label{Cij}
{C}_{ij}^{\rm (vc)}(i,j=e,d,s,b) ~=~ 
 - \frac{\tan ^{2}\beta }{2m_{A}^{2}} \, \frac{
Y_{i}^{SM}Y_{j}^{SM}\sin (\delta _{i}-\delta _{j})}{|1+\Delta m_i/m_i^{(0)}
||1+\Delta m_j/m_j^{(0)} |}\,,
\ee
where the phases $\delta_i$ are given in the expressions (\ref{phdiff}), 
(\ref{mfull}) and (\ref{Hfull}).
Eq.~(\ref{Cij}) shows that the crucial step in calculating the 
CP-odd operators $C_{ij}$ is finding the one-loop correction to the 
masses of the $d,s,b$ quarks and the electron.

%%%%%%%%%%%%%%%%%%%%%%%
\begin{figure}
 \centerline{%
   \psfig{file=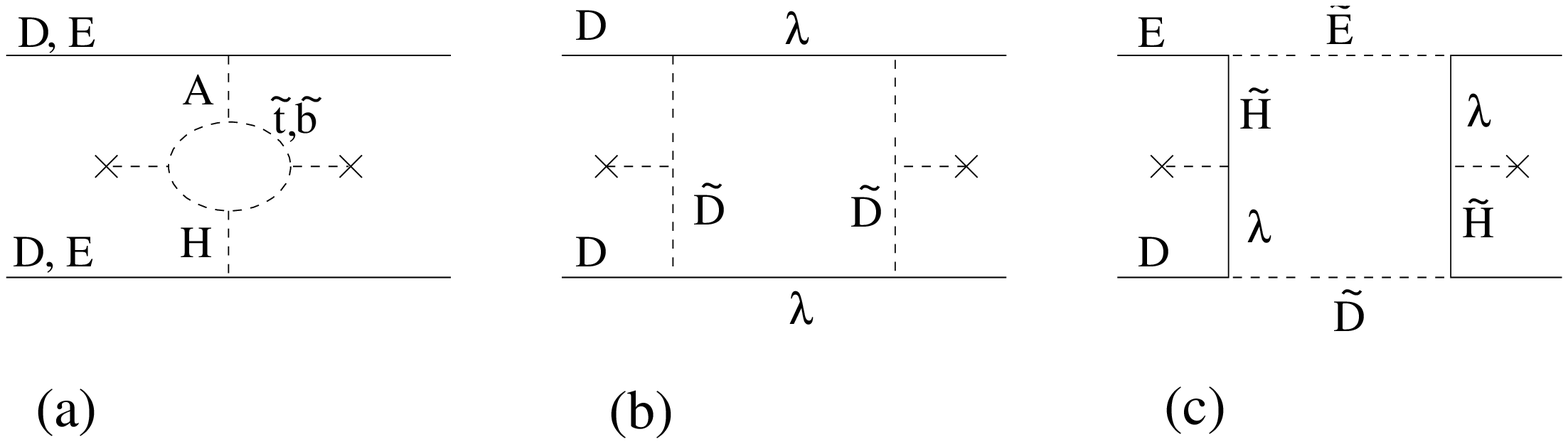,width=14cm,angle=0}%
         }
\vspace{0.1in}
 \newcaption{\small Diagrams that generate 
subleading contributions to $C_{ij}$. 
  (a) is the radiatively generated $H-A$ mixing, 
 and (b)-(c) are typical box diagrams.}
\end{figure}
%%%%%%%%%%%%%%%%%%%%%%%

At subleading order in $\tan\beta$, there are 
a large number of diagrams that contribute to $C_{ij}$, examples 
of which are given in Fig.~3. 
The complete calculation of these diagrams goes 
beyond the scope of the present paper. However, as shown in Ref. \cite{LP}, 
the CP-odd mixing of $H-A$ Higgs bosons 
may sometimes be numerically important and comparable 
to (\ref{Cij}). However, since the $H-A$ mixing term contains either
a loop factor which is not compensated by large $\tan\beta$ or a
suppression by $(Y^{SM}_{b(\tau)})^2\tan\beta$, $H-A$ mixing 
may  only become significant with the help of an additional mass 
hierarchy. %\footnote{The $\langle AH \rangle $--generated
%$C_{ij}^{\rm (AH)}$ grow faster than $C_{ij}^{\rm (vc)}$ with $\tan\beta$,
%thus becoming very significant for large 
%$\tan\beta$ ($>50$).}. 
In particular, 
if the following combination is large \cite{LP,Pil02},
\be
 \frac{A^2\mu^2 v^2}{m_{\rm squark}^4 m_A^2} \gg 1,
\label{condHA}
\ee
$H-A$ mixing may provide a sizeable contribution to the EDMs. 
The box diagrams scale differently
and lose their importance when the superpartners
are heavy, 
$m_0 \sim m_{1/2} \gg v$. Since this is one of the 
most interesting domains of the parameter space, we  
include  the effects of $H-A$ mixing and neglect the box diagrams. 

The coefficients $C_{ij}$ induced by diagrams 3a can be calculated 
with the use of the following formula:
\ba
C_{ij}^{(AH)}\simeq \frac{\langle AH\rangle\tan ^{2}\beta }{2m_{A}^{4}}
\,\frac{
Y_{i}^{SM}Y_{j}^{SM}}
{|1+\Delta m_i/m_i^{(0)}||1+\Delta m_j/m_j^{(0)} |},
\label{AH}
\ea
where $i,j = d,s,b,e$ and 
$\langle AH\rangle$ stands for the one-loop--generated amplitude in the 
effective Higgs Lagrangian,
\be
{\cal L}_{H,A} = - \fr{1}{2} m_A^2 A^2 - \fr{1}{2} m_H^2 H^2
-\langle AH\rangle AH + \cdots,
\label{effHA}
\ee
where the $\langle AH \rangle$ mixing parameter has the dimension of 
mass squared. Eq.~(\ref{AH}) 
also uses $m_{H}^{2}\simeq m_{A}^{2}\gg \langle AH\rangle$ and 
$\tan\beta\gg1$. The effect of $A-h$ mixing, where 
$h \sim  {\rm Re} \, (H_2 - \langle H_2\rangle ) $, 
is negligible under these conditions. 
In this approximation, it is not necessary to re-diagonalize the 3 $\times$ 3
Higgs mass matrix. It suffices to simply account for $\langle AH \rangle$ by 
a direct insertion on the Higgs line in Fig. 3a.
Note that $C_{ij}^{(AH)}$ 
are flavor-symmetric,
\be
C_{ij}^{(AH)} =C_{ji}^{(AH)},
\ee
and, unlike the vertex corrections, can generate a single flavor
four-fermion operator.

%{\em Comment: in a theory without the decoupling of 
%squarks and sleptons, indeed there 
%are quite a few box diagrams and there is no deep reason to prefer 
%Higgs exchange with 
%HA mixing over these box diagrams. So, our discussion of $\tan\beta^2$ 
%effects is 
%incomplete, unless we want to calculate all these disgusting boxes  }

\subsection{SUSY threshold corrections in the 
mass sector and scalar-pseudoscalar mixing}

In this subsection we provide a complete set of threshold corrections 
$\Delta m_i$, at leading order in $\tan\beta$, 
to the quark and lepton masses
(ignoring flavor mixing). 
Let us first briefly summarize our notation which, unless noted below,
follows that of Ref.~\cite{Haber:1984rc}. 
The chargino and neutralino mass matrices are diagonalized by
\begin{eqnarray}
&& U^* M_{\chi^+} V^\dagger = {\rm diag}(m_{\chi^+_1}, m_{\chi^+_2})\;, 
\nonumber\\
&& N^* M_{\chi^0} N^\dagger = {\rm diag}(m_{\chi^0_1},..., m_{\chi^0_4})\;,
\label{massmatrix}
\end{eqnarray}
where $U,V,$ and $N$ are unitary matrices. In addition, the 
sfermion mass matrix is diagonalized by the matrix
\begin{equation}
\left(  \matrix{c_f & s_f e^{i\phi_f} \cr
                -s_fe^{-i\phi_f} & c_f}    \right)\,,
\end{equation}
that rotates the basis $(\tilde f_L,~\tilde f_R)$ to the 
mass eigenstate basis $(\tilde f_1,~\tilde f_2)$.
Here $c_f \equiv \cos \theta_f$, $s_f \equiv \sin \theta_f$, and $\theta_f$ is 
the sfermion mixing angle (see \cite{Haber:1984rc}  for the explicit form 
of the mass matrices). In our convention the CP-violating phase  
appearing in the sfermion mixing mass is given by\footnote{This convention for
the phase of $A$ corresponds to the following normalization of the left-right
mixing terms in the squark mass matrix: $m_{LR}^2 = -m_f(A_f^* +\mu R_f)$.}
\begin{equation}
\phi_f = {\rm Arg}\left[-(A_f^* +\mu R_f)\right]\;,
\end{equation}
where $R_f=\tan\beta$ for $I_3=-1/2$ and $R_f=\cot\beta$ for $I_3=1/2$.

%Supersymmetric loops induce a correction to the mass term of a fermion $f$:
%\begin{equation}
%{\cal L}=-(m_f + \Delta m_f) f_R^\dagger f_L + {\rm h.c.} \;,
%\end{equation}
%where $m_f$ is the tree level mass term.  This loop--corrected mass 
%term can be made real by a phase redefinition of the right--handed field
%\begin{equation}
%f_R = f_R' ~e^{i \phi_m} \;\;,\;\; \phi_m={\rm Arg}(1+ \Delta m_f/m_f) \;.
%\end{equation}
%The phase $\phi$ then appears in the Higgs interactions. 

As argued in the previous section, a large value of $\tan \beta$ can 
compensate the loop suppression of $\Delta m_i$ 
such that the consequent CP-violating  phase is of order one. 
Below we list the relevant supersymmetric contributions
to $\Delta m_i$ which grow linearly with  $\tan\beta$. 
We omit the superscript in 
$m_i^{(0)}$ for brevity.  

\bigskip
\noindent{\bf  (i) Corrections to the down--type quark masses}

\noindent For all three down-type flavors, $d$, $s$ and $b$ (where 
$m_3$ and $\phi_3$ denote the gluino mass and its phase, and 
$\alpha_{_W}=g^2/ 4 \pi$): 
\begin{eqnarray}
\left(  \Delta m_d  \right)_{\lambda_3}&=& {2 \alpha_s \over 3 \pi} e^{-i(\phi_d+ \phi_3)}
s_d c_d m_3 (m_{\tilde d_1}^2 - m_{\tilde d_2}^2) I(m^2_{\tilde d_1},m^2_{\tilde d_2}, m^2_3)\;,
 \nonumber\\
\left(  \Delta m_d  \right)_{\chi^+}&=& - {\alpha_{_W} m_d \over  4\sqrt{2}\pi  m_{_W} \cos\beta}
\sum_i U_{i2}^* V_{i1}^* m_{\chi^+_i} \left[  c_u^2 \tilde B_0(m^2_{\tilde u_1},m^2_{\chi^+_i})      + s_u^2 \tilde B_0(m^2_{\tilde u_2},m^2_{\chi^+_i})  \right] \;, \nonumber \\
\left(  \Delta m_d  \right)_{\chi^0}&=& -{\alpha_{_W} \over 36 \pi} e^{-i\phi_d}
\tan \theta_W  s_d c_d (m_{\tilde d_1}^2 - m_{\tilde d_2}^2)
\sum_i N_{i1}^* (N_{i1}^* \tan\theta_W -3 N_{i2}^*) 
m_{\chi^0_i} \nonumber\\
&\times& I(m^2_{\tilde d_1},m^2_{\tilde d_2}, m^2_{\chi^0_i}) \nonumber\\
&+& {\alpha_{_W} m_d \over 24 \pi m_{_W} \cos\beta} \sum_i N_{i3}^* (N_{i1}^* \tan\theta_W - 3 N_{i2}^*)
m_{\chi^0_i} 
 \Bigl[  c_d^2 \tilde B_0(m^2_{\tilde d_1},m^2_{\chi^0_i})     \nonumber\\ 
&+& s_d^2 \tilde B_0(m^2_{\tilde d_2},m^2_{\chi^0_i})    \Bigr]  \\
&-& {\alpha_{_W} m_d \over 12\pi m_{_W} \cos\beta} \tan\theta_W \sum_i N_{i3}^* N_{i1}^* 
m_{\chi^0_i} 
 \left[  s_d^2 \tilde B_0(m^2_{\tilde d_1},m^2_{\chi^0_i})      + c_d^2 \tilde B_0(m^2_{\tilde d_2},m^2_{\chi^0_i})    \right] .\nonumber
\end{eqnarray}
\noindent In addition, for the $b$-quark there is a ``pure higgsino'' contribution:
\begin{eqnarray}
\left(  \Delta m_b  \right)_{\chi^+} &=&
-{\alpha_{_W} m_t m_b \over 8\pi m_{_W}^2 \sin\beta \cos\beta} e^{i \phi_t} 
c_t s_t (m_{\tilde t_1}^2 - m_{\tilde t_2}^2) \sum_i
U_{i2}^* V_{i2}^*  m_{\chi^+_i} \nonumber\\
&\times& I(m^2_{\tilde t_1},m^2_{\tilde t_2}, m^2_{\chi^+_i}) \;. 
\end{eqnarray}

\bigskip
\noindent{\bf (ii) Corrections to the charged lepton masses}
\begin{eqnarray}
\left(  \Delta m_e  \right)_{\chi^+} &=&
-{\alpha_{_W} m_e \over  4\sqrt{2}\pi m_{_W} \cos \beta } \sum_i U_{i2}^* V_{i1}^* m_{\chi^+_i}
\tilde B_0(m^2_{\tilde \nu_L}, m^2_{\chi^+_i} ) \;,  \\
\left(  \Delta m_e  \right)_{\chi^0} &=&
{\alpha_{_W} \over 4\pi } e^{-i \phi_e}  \tan\theta_W  
s_e c_e (m_{\tilde e_1}^2 - m_{\tilde e_2}^2) \sum_i 
N_{i1}^* (N_{i2}^* + \tan\theta_W N_{i1}^* ) 
 m_{\chi^0_i} \nonumber \\
&\times& I(m^2_{\tilde e_1},m^2_{\tilde e_2}, m^2_{\chi^0_i}) \nonumber\\
&-& {\alpha_{_W} m_e \over 8\pi m_{_W} \cos\beta } \sum_i N_{i3}^* (N_{i1}^* \tan\theta_W + N_{i2}^*)
m_{\chi^0_i}   
\Bigl[  c_e^2 \tilde B_0(m^2_{\tilde e_1},m^2_{\chi^0_i})     \nonumber\\ 
&+& s_e^2 \tilde B_0(m^2_{\tilde e_2},m^2_{\chi^0_i})    \Bigr]  \\
&-& {\alpha_{_W} m_e \over 4\pi  m_{_W} \cos\beta} \tan\theta_W \sum_i 
N_{i3}^* N_{i1}^* m_{\chi^0_i} 
\left[  s_e^2 \tilde B_0(m^2_{\tilde e_1},m^2_{\chi^0_i})      + c_e^2 \tilde B_0(m^2_{\tilde e_2},m^2_{\chi^0_i})    \right] . \nonumber
\end{eqnarray}

The two loop functions entering these expressions, $I$ and $\tilde B_0$, 
are given by
\begin{eqnarray}
&& I(x,y,z)= {x y \ln x/y + y z \ln y/z +x z \ln z/x \over
(x-y)(y-z)(x-z)} \;, \nonumber\\
&& \tilde B_0(x,y)= {x \ln x - y \ln y  \over x - y  } \;, 
\end{eqnarray}
where the tilde in $\tilde B_0$ serves to distinguish this ``truncated'' 
version\footnote{Note that the expression for $\tilde B_0$
requires a regulator scale.  However, our physical results do 
not depend on this scale.} of the Passarino--Veltman function
\cite{Passarino:1978jh} (following the definitions of 
\cite{Lebedev:2001ez}). 

Let us briefly discuss these results.
Firstly, to gain some intuition on the 
relative contributions, it is instructive 
to consider a limiting case,
where the left--right mixing is considered a perturbation.
The dominant corrections arise from $(\De m_{(d,s,b)})_{\lambda_3}$ and
$(\De m_b)_{\ch^+}$. For the first of these, taking 
$m_3\sim m_{\tilde{d}_1} \sim m_{\tilde{d}_2}$, we have (for $d=(d,s,b)$)
\be
 \left(\frac{\De m_{d}}{m_{d}}\right)_{\lambda_3} 
 \sim \, \frac{\al_s}{3\pi} ~
  \frac{m_3}{m_{\tilde{d}}^2}\,(\mu \tan\beta +A^*). \label{delm_s}
\ee
The contribution of these threshold corrections to $\de_d$ arises 
only from Im$(\mu)$, as the contribution from Im$(A)$ is cancelled
in constructing $\de_d^{(m)} - \de_d^{(H)}$. From (\ref{delm_s}),
we see that when $\mu$ is negative ($\th_\mu=\pi$) there is destructive
interference with the tree-level mass term.

The presence of an extra diagram sets the $b$ flavor 
apart from $d$ and $s$, for which the following relation is expected if 
the soft-breaking sector is flavor-blind:
\be
\label{d=s}
\frac{\Delta m_d}{m_d}= \frac{\Delta m_s}{m_s}, 
\;\;\;{\rm and}\;\;\; \delta_d = \delta_s.
\ee
For the $b$ quark mass, in a regime where the gaugino--higgsino mixing  
is also small, namely $\mu, m_i \gg m_{_W}$ with $m_i$ being  the gaugino masses,
we also find
\be
 \left(\frac{\De m_b}{m_b}\right)_{\ch^+} \sim \, - \frac{(Y_t^{SM})^2 }{32\pi^2} ~
   \frac{\mu^*A_t^*}{m_{\tilde{t}}^2}\tan\beta ,
\ee
which indicates that when $\mu$ is real and positive, so that 
$(\De m_{(d,s,b)})_{\lambda_3}$ is real, the contribution to $\de_b$ is 
positive when $0< \th_A < \pi$.

As is apparent from the limiting cases above, we 
observe the appearance of the reparametrization invariant CP-violating phases.
In particular, at large $\tan\beta$
\begin{equation}
\phi_d ~\longrightarrow~  \theta_\mu  + \pi \;,
\end{equation}
where $\th_\mu \equiv {\rm Arg}\,(\mu)$, such that the CP-odd phase entering 
the gluino expression is Arg$\,(\mu m_3)$. The analogous behavior
of the neutralino contributions is seen most easily in the 
limit $\mu, m_i \gg m_{_W}$. Then,
\begin{equation} 
N_{i1}^* ~\longrightarrow~ \delta_{i1}~ e^{-{i\over 2} {\rm Arg} ~m_1},
\end{equation}
and the consequent CP-odd phase in the gaugino piece of 
the neutralino contribution is Arg$\,(\mu m_1)$ 
(or Arg$\,(\mu \sqrt{m_1 m_2})$). For the higgsino contribution, the 
relevant phase is Arg$\,(A\mu) \equiv \theta_A +\theta_\mu$.

It is worth noting that the origin of the $\tan\beta$--enhancement, 
which is not manifest in the general expressions. For the gaugino 
contributions proportional to the $I$-function, this arises 
implicitly via the relation 
\begin{equation}
m_{\tilde d_1}^2 - m_{\tilde d_2}^2 ~\propto~ m_d \tan\beta 
\end{equation}
at large $\tan\beta$, whereas the mixing angle stays constant in this limit.

The $\tan\beta$ enhancement of 
the amplitudes involving the $\tilde B_0$ function
deserves special discussion.
The relevant diagrams  in the mass eigenstate basis  appear to be divergent
since they are proportional to the Passarino--Veltman $B_0$ function. However their 
$\tan\beta$--enhanced imaginary parts are finite.
As an example, let us consider the chargino contribution which involves
gaugino--higgsino mixing. The relevant expression appearing in the 
amplitude is
\begin{equation} 
{\cal G}= {1\over \cos\beta}  \sum_i U_{i2}^* V_{i1}^* m_{\chi^+_i}  \tilde B_0(m^2_{\tilde u_1},m^2_{\chi^+_i}) \;.
\end{equation}
There are in fact several compensating factors of $\tan\beta$ in
this expression. To see this, we note that   
Eq.(\ref{massmatrix}) implies
\begin{equation}
U^* f \left[ M_{\chi^+}  M_{\chi^+}^\dagger \right] M_{\chi^+} V^\dagger = 
f \left[{\rm diag}(m_{\chi^+_1}^2, m_{\chi^+_2}^2)  \right]
{\rm diag}(m_{\chi^+_1}, m_{\chi^+_2})
\end{equation}
for any (analytic) function $f$. Therefore,
\begin{equation}
{\cal G}= {1\over \cos\beta} \left[   \tilde B_0 \left( m^2_{\tilde u_1}, M_{\chi^+}  M_{\chi^+}^\dagger \right) M_{\chi^+}
\right]^*_{21} \;\;.
\end{equation}
If $\tilde B_0$ here is replaced by a constant independent of $M_{\chi^+}$,
no $\tan\beta$ enhancement appears since
$\left[ M_{\chi^+} \right]_{21}= \sqrt{2} m_{_W} \cos\beta$.
This is the reason why the $\tan\beta$--enhanced contribution is finite and 
the original Passarino--Veltman function gets
replaced by its ``truncated'' version. 

It is instructive to expand $\tilde B_0$ in powers of 
$(M_{\chi^+}  M_{\chi^+}^\dagger - m^2_{\tilde u_1})/m^2_{\tilde u_1}$
and extract the first $\tan\beta$-enhanced term.
Using the explicit form of the chargino mass matrix, one finds
%\begin{eqnarray}
%&&\tilde B_0\left( m^2_{\tilde u_1},M_{\chi^+}  M_{\chi^+}^\dagger\right)
% \longrightarrow  
%{M_{\chi^+}  M_{\chi^+}^\dagger - m^2_{\tilde u_1} \over 2 m^2_{\tilde u_1}}
%\;\; +\;\;  \left( M_{\chi^+}  {\rm - independent\;terms } \right) \nonumber
%\end{eqnarray}
%and 
%\begin{equation}
%\left[  M_{\chi^+}  M_{\chi^+}^\dagger  M_{\chi^+}   \right]_{21}=
%\sqrt{2} m_{_W} \mu m_2  ~ \sin\beta   \;\;+\;\; ( \cos\beta-{\rm terms})\;,
%\end{equation}
%so 
\begin{equation}
{\cal G} \longrightarrow  {\mu^* m_2^* \over \sqrt{2} m^2_{\tilde u_1} } m_{_W} \tan\beta \;.
\end{equation}
This result can be understood via the mass--insertion 
approximation \cite{Blazek:1995nv,Pierce:1996zz}. To obtain
$\tan\beta$ enhancement, it is necessary to introduce three 
mass insertions on the chargino line:
$\mu$ -- to mix $\tilde H_1$ with $\tilde H_2$, $m_{_W} \sin\beta$ -- to mix
the higgsinos and gauginos, and $m_2$ -- to get 
the chirality flip in the diagram. Similar considerations apply 
to the neutralino contributions.
We note that analogous calculations have been performed in
Refs.\cite{Pierce:1996zz} and \cite{Ibrahim:2003ca}.

{\bf (iii) Scalar-pseudoscalar mixing}

The computation of $H-A$ mixing is straightforward, especially in the limit 
$m_A^2\gg \langle AH\rangle$ \cite{HA}. Non-negligible contributions 
arise only from stop, sbottom and stau loops:
\ba
\nonumber\langle AH\rangle &=& \frac{3}{8\pi^2 v^2}
\left[\fr{m_t^4|\mu|^2|A_t|^2}{(m^2_{\tilde t_1} -m^2_{\tilde t_2})^2}
E(m^2_{\tilde t_2}/m^2_{\tilde t_1})
 \sin (2\theta_{A_t} +2 \theta_\mu) \right.\\ &&+\left.
\fr{m_b^4|\mu|^2|A_b|^2\tan^4\beta  }{(m^2_{\tilde b_1} -m^2_{\tilde b_2})^2}
E(m^2_{\tilde b_2}/m^2_{\tilde b_1})
 \sin (2\theta_{A_\tau} +2 \theta_\mu)\right.\\ &&+\left.\nonumber
\fr{m_\tau^4|\mu|^2|A_\tau|^2\tan^4 \beta }{3(m^2_{\tilde \tau_1} -m^2_{\tilde \tau_2})^2}
E(m^2_{\tilde \tau_2}/m^2_{\tilde \tau_1})
 \sin (2\theta_{A_\tau} +2 \theta_\mu)
\right],
\ea
where the loop function is given by $E(a) = - 2 + (a+1)(a-1)^{-1}\ln a$.
At moderate $\tan\beta$ the stau and 
sbottom contributions are heavily suppressed 
due to the factor of $m_{b(\tau)}^4$ in the numerator. However, this 
is not true at large $\tan\beta$, as  the fourth power of 
$\tan\beta$ can overcome this suppression.

It is important to note that $\langle AH\rangle$ is even under
flipping the sign of $A$, unlike the effects
of the vertex corrections. Thus, these two sources can interfere
$constructively$ or $destructively$, depending on the position in
parameter space. For example, choosing $\th_\mu = 0$ and $0<\th_A<\pi/2$
leads to destructive interference between 
$C_{ij}^{\rm (vc)}$ and $C_{ij}^{(AH)}$.

\section{Synopsis of the EDM formulae}

In this section, we compile  the different contributions into physical 
observables measured in various EDM experiments. Such observables can be 
subdivided into three main categories: EDMs of paramagnetic 
atoms and molecules, EDMs of diamagnetic atoms, and the neutron EDM.

\subsection{EDMs of paramagnetic atoms -- thallium EDM}

Among various paramagnetic systems, the EDM of 
the thallium atom currently provides
the best constraints on fundamental CP violation and is often 
interpreted directly in terms of the 
EDM of the electron. We therefore specialize our analysis 
to the case of Tl, noting that the conditions making  $C_{ij}$ important 
for thallium will also be applicable to other paramagnetic systems such as 
the YbF and PbO molecules,  where significant experimental 
improvements in EDM measurements are likely \cite{YbF,PbO}.

The semileptonic four-fermion operators (\ref{4f}) 
induce the following T-odd nucleon-electron 
interactions\footnote{Note that we differ from Ref.~\cite{KL} in 
the definition of $\gamma_5$, so that we use $P_L=(1-\gamma_5)/2$.}
\cite{KL},
\be
{\cal L}_{eN} = C_S \bar e i \gamma_5 e \bar NN +  C_P \bar e e \bar N i\gamma_5 N
+ C_T \epsilon_{\mu\nu\alpha\beta}
\bar e\sigma^{\mu\nu} e \bar N \sigma^{\alpha\beta} N,
\label{eN}
\ee
where the isospin dependence is suppressed. 
Among these couplings, $C_S$ plays by far the most important 
role for the EDM of paramagnetic 
atoms because it couples to the spin of the electron and is enhanced by the
large number of nucleons in heavy atoms. 

A number of atomic calculations \cite{LiuKelly,MP1,MP2} 
(see also Ref. \cite{KL} for a more complete list) have 
established the relation between the EDM of 
thallium, $d_e$, and the coefficients of the 
CP-odd electron-nucleon interactions $C_S$:
\be 
d_{\rm Tl} = -585 d_e -   e\ 43 ~{\rm GeV}
 \times ( C_S^{sing} - 0.2 C_S^{trip}).
\label{dtl}
\ee
The relevant atomic matrix elements are calculated to within 
$10-20\%$ accuracy \cite{fgreview}. Here the relative contribution of the 
isospin-triplet coupling is suppressed by $(N-Z)/A \simeq 0.2$, 
and the effects of $C_P$ and $C_T$ are negligible. 

Before turning to the contributions to $C_S$, we recall for completeness
(and to fix our conventions)
that the electron EDM $d_e$ receives a number of well-known contributions from
superpartner loops. At one-loop order, threshold corrections from
$\tilde{e}-\ch^+$ and  $\tilde{e}-\ch^0$ loops dominate, and we follow
the notation of Ibrahim and Nath \cite{IN} for these 
terms (up to a different convention for the squark mixing
angles as specified earlier). At two-loop
order, non-negligible Barr-Zee--type Higgs-mediated graphs 
contribute \cite{ckp},
where CP violation enters through the couplings of (s)fermions to 
the pseudoscalar Higgs $A$. We follow the work of Chang, Keung and
Pilaftsis \cite{ckp} (though our phase conventions are somewhat different)
and to emphasize an additional contribution, we quote the full
expression below:
\be
 d_e^{\rm two-loop} = |e| Q_e \frac{\al_{em}}{32\pi^3} \frac{m_e}{m_A^2}
   \sum_{j=t,b,\tau} \zeta_j Q_j^2 N_j
   \left[ F\left(\frac{m_{\tilde{j}_1}^2}{m_A^2}\right)
  - F\left(\frac{m_{\tilde{j}_2}^2}{m_A^2}\right)\right]\tan\beta,
\ee
where $j=t,b,\tau$ are flavors running in the sfermion loop. $N_t=N_b=3$, 
and $N_\tau=1$, accounts for the trace over color,
while the two-loop function $F(z)$ is given explicitly in \cite{ckp}.
The CP-violating couplings in our conventions are:
\be
  \zeta_j = \frac{\sin(2\th_j)m_jR_j{\rm Im}(\mu e^{-i\ph_j})}
    {v^2\sin\beta\cos\beta},
\ee
Note that while there are in principle a large 
number of possible diagrams in this
class \cite{pilaf_2}, those containing stop, sbottom, and stau loops should
be dominant at large $\tan\beta$. We emphasize here that the stau-loop
contribution, which has hitherto been neglected in the literature, is 
a significant addition -- in fact for the CMSSM and NUHM at $\tan\beta =50$ 
to be considered here, it often provides the 
dominant contribution due to the generically light staus.

In order to see the dependence on the reparametrization invariant
phases, it is useful to
again consider the small-mixing regime where 
$m_L^2\sim m_R^2 \gg m_A^2,m_{LR}^2$. For large $\tan\beta$, the
EDM then reduces to
\be
 d_e^{\rm two-loop}(j=t,b,\tau) \sim - Q_j^2 N_j (Y^{SM}_j)^2 
 \frac{\al_{em} m_e |\mu A_j|}{192\pi^3 m_{\tilde{j}}^4} 
 \ln \left(\frac{m_{\tilde{j}}^2}{m_A^2}\right)
 \sin(\th_{A_j} +\th_\mu)\tan^n\beta,
\ee
where we used $Q_e = -1$, and defined $n$ as $n=1$ for $j=t$ and $n=3$ 
for $j=b,\tau$.
It is important to recognize that when $\th_\mu=0$ the two-loop contribution to 
the electron EDM is
negative which interferes destructively with the one-loop contribution
referred to above\footnote{
Note that Barr-Zee--type diagrams with chargino loops also contribute for 
$\th_\mu\neq 0$ \cite{Chang:2002ex,Pil02}.} . 

Having accounted for $d_e$, 
the calculation of $C_S$ induced by (\ref{4f}) follows 
from the low-energy theorem that 
expresses the nucleon matrix element of 
$\alpha_s G_{\mu\nu} G^{\mu\nu}$ in terms of the mass of the 
nucleon and the QCD beta function \cite{SVZ}. As a consequence 
of this theorem, for a heavy quark $Q$, 
$\langle N| m_Q \bar QQ |N\rangle \simeq 70 $ MeV. 
This theorem is valid in the exact chiral limit, and thus receives 
corrections from the non-zero values of the light quark masses. 
 Following \cite{LP}, we parametrize the $m_s\bar ss$ matrix element  as 
$\kappa \equiv \langle N|m_{s}\overline{s}s|N\rangle /220$ MeV. 
Using this parametrization, as well as the measured value of 
$(m_{u}+m_{d})\langle N|\overline{u}u+\overline{d}d|N\rangle /2\simeq 45$ MeV 
and the ratio of the quark 
masses $m_u/m_d \simeq 0.55$, we arrive at the following 
general formula for $C_S$ in terms of $C_{qe}$,
\begin{eqnarray}
\label{sintrip}
C_S^{sing} &=& C_{de}\frac{29~{\rm MeV}}{m_d}+ C_{se}\frac{\kappa \times
220~{\rm MeV}  }{m_s}+ 
C_{be}\frac{66~{\rm MeV}(1-0.25 \kappa )}{m_b},
\\
C_S^{trip} &\sim&- C_{de}\frac{1-3~{\rm MeV}}{m_d}.
\nonumber
\end{eqnarray}
In this formula, $C_{de}/m_d$ and $C_{se}/m_s$ should be 
taken at the normalization scale  
1 GeV, while $C_{be}/m_b$ must be normalized at $m_b$. 
In the case of minimal SUSY models,
these combinations are scale invariant, as $C_{qe} \sim m_q$. There is a rather 
large uncertainty in the 
isospin triplet coupling due to the poorly 
known value of $(m_d-m_u)\langle p|\bar uu 
- \bar dd | p\rangle $. Fortunately, $C_S^{trip}$ is numerically small
and a further suppression via (\ref{dtl}) makes the 
contribution of $C_S^{trip}$ completely negligible for all applications. 
Thus, from now on, we neglect $C_S^{trip}$ and drop the superscript 
in $C_S^{sing}$.

The largest uncertainty in $C_S$ (\ref{sintrip}) that may significantly 
affect numerical values of EDMs originates from the 
poorly known value of $\kappa$. 
The assumption that the $s$ quark behaves as a heavy quark 
would give $\kappa \simeq 0.3$, which we regard as a lower bound.
The analysis of the baryon octet mass splitting at leading order in
 chiral perturbation theory instead suggests $\kappa \simeq 1$, 
while an improved next-to-leading order calculation 
gives a prediction $\kappa \simeq 0.50 \pm 0.25$ \cite{KBM} that we use as the 
central point for our numerical analysis. From these numbers it 
is clear that the overall 
precision with which this matrix element can be estimated is 
not better than $30\%$. 

Substituting Eqs.~(\ref{Cij}) and (\ref{AH}) into (\ref{sintrip}), 
we arrive at the final expression for $C_S$
in terms of  $m_A$, $\Delta m_i$, $\delta_i$ and $\langle AH\rangle $: 
\ba
C_{S}\simeq - \frac{5.5\times 10^{-10}\tan ^{2}\beta }
{m_{A}^{2}|1+\De m_e/m_e^{(0)} |} 
 \!\!\!\!\!\!\!&&\left[ \frac{(1-0.25\kappa )}{
|1+\De m_b/m_b^{(0)} |}\left (\sin (\delta _{b}-\delta _{e})-
\fr{\langle AH\rangle}{m_A^2}\right)\right. \nonumber\\
&&+\frac{3.3\kappa}{|1+\De m_s/m_s^{(0)}|} 
\left(\sin (\delta _{s}-\delta _{e})-
\fr{\langle AH\rangle}{m_A^2}\right)\label{masterf}
\label{cstl}
\\
 &&+\left.\frac{0.5}{|1+\De m_d/m_d^{(0)}|}\left(\sin (\delta _{d}- 
\delta _{e})-
\fr{\langle AH\rangle}{m_A^2}\right)
\right] \;.  
\nonumber
\ea
We use this calculation of $C_S$, along with the standard 
SUSY calculations of $d_e$ reviewed above, in the analysis of the thallium EDM 
as a function of SUSY masses and phases which will be discussed in Section~5.

\subsection{Neutron EDM}

The calculation of the neutron EDM in terms of the Wilson coefficients of 
Eq.~(\ref{largetanbeta}) represents a difficult non-perturbative problem. 
The most common approach invokes ``naive dimensional analysis'' 
(NDA) \cite{GM,W} which gives an order of magnitude estimate 
for $d_n(d_i,\tilde d_i, w)$ with uncertain signs. 
Some partial calculations have also been performed with 
the use of chiral perturbation 
theory \cite{CDVW,KK2} and lattice QCD \cite{Aoki}. Recently, a universal 
treatment of all operators within the same method was developed 
in Refs.~\cite{PR,DPR} using QCD sum rule techniques. 
We briefly summarize these results below.

It is natural to expect that the dominant contribution 
to the neutron EDM comes from the 
EDMs and color EDMs of $u$ and $d$ quarks. We recall that the 
$\bar\theta$-dependence is removed by the Peccei-Quinn mechanism,
which at the same time reduces the contribution of the color EDM 
of the $s$-quark \cite{PR}. The QCD sum rule analysis \cite{PR} 
leads to the following result:
\be
 d_{n}(d_i, \tilde d_q) = 0.7(d_d-0.25d_u) + 0.55e(\tilde d_d + 0.5\tilde d_u),
\label{dn1}
\ee
where the value of the quark vacuum condensate 
$\langle \bar qq\rangle = (225 \, {\rm MeV})^3$ has been used. 
Here $\tilde d_q$ and $d_q$ are to be normalized at the hadronic scale 
which we assume  to be 1 GeV. The relation to the Wilson coefficients 
at the weak scale is as follows,
\be
\tilde d_q(1\,{\rm GeV}) \simeq 0.91 \tilde d_q(M_Z);~~~
d_q(1\,{\rm GeV}) \simeq 1.2 d_q(M_Z).
\ee
Note that $\tilde d_q$ as defined by Eq.~(\ref{leff}) has 
a much milder QCD scaling than an alternative definition
for $\tilde d_q$
which includes a factor of $g_s$ in its definition.
The reader should also note that the 
quark masses used for the SUSY calculations of $d_q$ and $\tilde d_q $ 
should be taken at the weak scale, where their numerical values are  smaller 
than the low energy values by a factor of $\sim 0.35$, {\em e.g.} 
$ m_d(M_Z) \simeq 9.5\,{\rm MeV} \times 0.35$.

The expression (\ref{dn1}) has several nice features. Its flavor composition 
agrees with the predictions of the SU(6) constituent quark model, 
while the proportionality to $d_q\langle \bar qq\rangle \sim 
 m_q\langle \bar qq\rangle \sim f_\pi^2m_\pi^2$ removes 
any sensitivity to the 
absolute value of the light quark mass.
The quark EDM part of (\ref{dn1}) is in  good agreement
with lattice calculations \cite{Aoki}. Finally, the 
overall uncertainty of (\ref{dn1}) is estimated  to be 
at the 30\%--50\% level \cite{PR}. 
For completeness, we recall that there are standard SUSY 
one- and two-loop contributions to $d_q$ and ${\tilde{d}}_q$ in 
analogy with those discussed for 
the electron EDM above. For 1-loop quark EDMs and CEDMs we follow
\cite{IN} and note that, in comparison to the electron EDM, for 
quarks there are additional one-loop 
diagrams containing squark-gluino loops. The 
two-loop Barr-Zee--type contributions to the quark EDMs \cite{ckp}
also receive a significant correction from the stau-loop in addition to
those containing stops and sbottoms.

The Weinberg operator can also be an important source of  CP violation, 
especially when the third generation of squarks is lighter than the first 
two generations. Unfortunately, a reliable calculation of its contribution 
to the neutron EDM is problematic at this point. The QCD sum 
rule approach \cite{DPR} produces an estimate that is close to the 
NDA result but this calculation is at the border--line of applicability of 
the method. With $w$ normalized at the 1 GeV scale one finds:
\be
  d_n(w) \sim 20{\rm MeV}\times e~w.
\label{dn2}
\ee
This estimate is assessed to be valid within a factor of 
2--3 \cite{DPR}. However, it seems unlikely that this uncertainty can
be significantly reduced. An important implication is then that
even the sign cannot be reliably inferred, as the next order 
corrections to the QCD sum rule determination (\ref{dn2})  
are not calculable and plausibly are large. 

When $\tan\beta$ is large, we expect that a generically dominant
contribution to the Weinberg operator will be induced by the $b$-quark 
color EDM (see also \cite{CKKY,ALN}). 
Thus, the expression for $w(\tilde d_b)$ linearly enhanced by
$\tan\beta$ takes the form: 
\be
w(1~{\rm GeV}) = 0.72 w(m_b) = 
- 0.72\times \fr{g_s^3\tilde d_b(m_b)}{32 \pi^2 m_b}
= - 0.68\times \fr{g_s^3\tilde d_b(M_Z)}{32 \pi^2 m_b}.
\label{wfromb}
\ee
It is well known that the $t$ and $c$ quark
two-loop contributions are just a fraction of the total EDM
at low $\tan\beta$ \cite{Dai:xh}. Since they do not grow with
$\tan\beta$, their effect is expected to be negligible but nonetheless 
they are included in our analysis. 
We remark at this point that it is common  in the SUSY-EDM literature to
quote the Weinberg operator at the hadronic scale as $3.3\times w(M_Z)$. 
This relation originates from Ref.~\cite{ALN} where the contributions to $w$ 
induced by $\tilde d_b$ and $\tilde d_c$
were added at the $b$ and $c$ thresholds with 
some specific assumptions about the
SUSY spectrum and a fixed (rather low) value of $\tan\beta$. 
For any other point in the parameter space this coefficient is going 
to be different, and therefore an explicit summation of the
$b$ and $c$ thresholds should be performed separately.

There are many different types of four-fermion operators generated via 
the vertex correction mechanism that contribute
to the neutron EDM at 
leading order: $C_{ds}$, $C_{sd}$, $C_{bd}$, $C_{db}$, $C_{bs}$, and $C_{sb}$. 
The contribution of the first two operators to $d_n$ could be important 
a priori \cite{PH}.
We note, however, that the CP-violating phases induced under renormalization  
are the same for the $d$ and $s$ quarks to very high accuracy within 
the framework
of a flavor-blind soft-breaking sector. 
Thus, $\delta_s \approx \delta_d$ 
and  $C_{d s} \approx -C_{s d} \approx 0$, from Eq.~(\ref{d=s}).
This however does not apply to $C_{bd}$ and $C_{db} $ or 
$C_{bs}$ and $C_{sb} $ due to an
additional chargino contribution to the $b$--quark mass. 
The estimates of $d_n$ induced by 
$C_{bd}$ and $C_{db}$ can be obtained along the lines of Ref.~\cite{DPR}. 
We supply some of the details of this calculation 
in Appendix B, and quote here the result:
\be
 d_n(C_{bd})\sim  
\fr{e~0.65\times 10^{-3} {\rm GeV}^2}{m_b}C_{bd},
\label{dn3}
\ee
where we took into account that $C_{bd}\simeq - C_{db}$. The 
Yukawa couplings $Y_d$ and $Y_b$ that enter the coefficient 
$C_{bd}$ in (\ref{dn3}) are normalized at 1 GeV and 
$m_b$ respectively. In the final result for $d_n$, $C_{bs}$ and $C_{sb}$
may be as important as $C_{bd}$ and $C_{db}$. However, we
are not aware of any reliable way to estimate $d_n(C_{bs}, C_{sb})$.

Combining the different 
contributions, (\ref{dn1}), (\ref{dn2}) and (\ref{dn3}),
we arrive at a final formula for the neutron EDM as a function of 
the different Wilson coefficients:
\be
d_n  = d_n(d_q, \tilde d_q) + d_n(w) + d_n(C_{bd}).
\label{dn_final}
\ee
One should keep in mind at this point that, as emphasized above,
Eqs. (\ref{dn2}) and (\ref{dn3}) are undoubtedly of poorer precision
than (\ref{dn1}). Therefore, a reliable calculation 
of $d_n$ in terms of a specific combination of SUSY CP-violating phases 
is possible only if both the second and third terms in (\ref{dn_final}) 
are smaller than  $d_n(d_q, \tilde d_q)$. When instead they 
contribute at a level comparable to the EDMs and color EDMs of quarks, 
an interpretation of the neutron EDM bound in terms of constraints
on specific CP-violating phases becomes problematic.

\subsection{EDMs of diamagnetic atoms -- mercury EDM}

EDMs of diamagnetic atoms, i.e. atoms with total 
electron angular momentum equal to zero, also provide an important test of CP 
violation \cite{KL}. 
The current limit on the EDM of mercury 
\cite{Hg} furnishes one of the most sensitive constraints on SUSY 
CP-violating phases \cite{FOPR}. However, the calculation of $d_{\rm Hg}$ is 
undoubtedly the most difficult as it requires 
QCD, nuclear, and also atomic input. 

The atomic EDM of mercury arises from several important sources, namely, 
the Schiff moment $S$ \cite{Schiff}, the electron EDM $d_e$, and also 
the electron-nucleus interactions $C_S$ and $C_P$ (see, e.g. Ref. \cite{KL}
for a comprehensive review). Schematically, the mercury EDM can 
be represented as
\be
d_{\rm Hg} = d_{\rm Hg}\left( S[\bar g_{\pi NN}(\tilde d_i, C_{q_1 q_2})], \, 
C_S[C_{qe}], \,
C_P[C_{eq}], \, d_e \right),
\label{dhgsch}
\ee
where $\bar{g}_{\pi NN}$ collectively denotes the 
CP-odd pion-nucleon couplings.  
The atomic and nuclear parts of the calculation have been performed 
by different groups, and several results 
such as $d_{\rm Hg}( S)$ \cite{Hgnew} and $S(\bar g_{\pi NN})$ \cite{DS}
have been updated recently. The most important numerical change comes from 
a new  QCD sum-rule calculation of $\bar{g}_{\pi NN}(\tilde d_i)$ \cite{P}, 
that obtained a  preferred range and ``best'' value for this  
coupling. Previous estimates \cite{KKY,FOPR} are within this 
preferred range, but close 
to the largest possible value for the $\bar{g}_{\pi NN}(\tilde d_i)$ 
matrix element. The best value that follows from the 
sum-rule analysis \cite{P} is a factor of $\sim 2.5-3$
smaller than previously used values \cite{FOPR}. Incorporating 
the changes in the atomic matrix elements, the current 
theoretical estimate for the mercury EDM induced by 
dimension 5 operators stands as: 
\be
d_{\rm Hg}= 7\times 10^{-3}e(\tilde d_u - \tilde d_d)  + 10^{-2}\times d_e.
\label{hglowtg}
\ee
Note that the electron EDM also contributes to $d_{\rm Hg}$, 
although less significantly than to $d_{\rm Tl}$ which provides
a more stringent bound. Therefore, the most valuable feature of
$d_{\rm Hg}$ is its sensitivity to the triplet combination of
color EDM operators $\tilde d_i$, which  surpasses the neutron EDM 
sensitivity to this combination of these 
operators by a factor of a few. The overall uncertainty
in the QCD part of the calculation is considerably 
larger than that for the neutron due to significant cancellations, 
reflecting the fact that the result vanishes in  the vacuum 
factorization approximation (see Appendix B). However, the
dominant dependence on the $(\tilde d_u - \tilde d_d)$ combination
ensures that these uncertainties enter as an overall factor
and therefore do not significantly alter the 
shape  of the unconstrained band of the parameter space in
the $\theta_\mu - \theta_A$ plane. 

One should note that the mercury EDM also 
receives contributions from the color EDM of the strange quark \cite{FOPR} 
and the EDMs of light quarks, although their 
contribution is subleading. Finally, the Weinberg operator does not provide 
any appreciable contribution to $d_{\rm Hg}$ because its 
contribution to $\bar{g}_{\pi NN}$
is suppressed by an additional chiral factor  of 
$m_q/1\,{\rm GeV}\sim 10^{-2}$. 

At large $\tan\beta$ one should also 
include additional contributions to the Schiff moment 
coming from the four-quark operators, as well 
as the effects of the semi-leptonic 
operators $C_S$ and $C_P$:
\begin{eqnarray}
d_{\rm Hg} &=& 7\times 10^{-3}\,e\,(\tilde d_u - \tilde d_d)  + 10^{-2}\, d_e
 \nonumber \\
\label{Hgmaster}
 && -1.4\times 10^{-5}\,e\,{\rm GeV}^2 
\left( \fr{0.5 C_{dd}}{m_d} +3.3 \kappa \fr{ C_{sd}}{m_s}+
\fr{C_{bd}}{m_b}(1-0.25\kappa)\right)
\\\nonumber
 && +3.5\times 10^{-3}{\rm GeV}\, e\, C_S \;+\;
4 \times  10^{-4}{\rm GeV}\, e\, C_P.
\end{eqnarray}
The second line in (\ref{Hgmaster}) is the contribution of the 
four-quark operators to the Schiff moment. The details of 
the $\bar{g}_{\pi NN}(C_{q_1q_2})$ estimates, based on a factorization 
assumption, are given in Appendix B. The result depends on the 
same parameter $\kappa$ (the nucleon strangeness in the $0^+$ channel)
as the coefficient $C_S$. As emphasized before, $C_{sd}$ and $C_{dd}$ are 
not induced via vertex corrections, and arise only through $A-H$ mixing. 
Finally, the contribution of $C_P$ is given primarily by $C_{ed}$ 
\cite{LP} while the contributions from $C_{es}$ and $C_{eb}$ 
are consistent with zero. The QCD normalization 
of all operators is the same as in the thallium and neutron cases. 

In the region of parameter space where the
color EDM contribution to $d_{\rm Hg}$ is dominant, the mercury EDM 
limit is extremely valuable since, as noted above, it constrains
the triplet combination of up and down color EDMs, and 
therefore a calculable 
combination of the CP-violating phases. 
On the other hand, when other terms in Eq.~(\ref{Hgmaster}) 
contribute at a level comparable to that of 
$\tilde d_u - \tilde d_d$, 
which in fact is expected at large $ \tan \beta$,
different 
uncertainties no longer factorize in front of a given phase 
combination, and the interpretation of the 
mercury EDM constraint becomes more difficult.

\section{Numerical analysis of EDMs at large \boldmath{$\tan\beta$}}

In this section we analyze all three  
observables, $d_{\rm Tl}$, $d_n$ and $d_{\rm Hg}$, 
within two classes of MSSM models. In both cases we assume  
flavor-blind SUSY breaking, 
a common trilinear soft-breaking parameter $A_0$, and
a universal (real)  gaugino mass parameter  $m_{1/2}$ 
at the GUT scale. 
This narrows down the number of SUSY  CP-violating phases to two, which
can be identified with the phases of the $\mu$ and $A_0$ parameters. 
We then perform an EDM analysis 
separately for these two phases. In all numerical runs the 
value of $\tan\beta$ 
was chosen to be nearly maximal, $\tan\beta = 50$. 
We defer an analysis of the constraints on the SUSY CP-violating phases
to a subsequent publication
\cite{DLOPR2} and now focus on the relative importance
of various EDM contributions.

The constrained MSSM (CMSSM) is a popular framework defined by the
following set of universal SUSY parameters at the GUT scale:
\begin{equation}
\left\{\tan\beta,~m_0 ,~ m_{1/2},~\vert A_0 \vert,~ \theta_A,~ 
  \theta_\mu\right\}.
\end{equation}
The magnitude of the $\mu$--parameter and the pseudoscalar mass
are determined by the radiative electroweak symmetry breaking conditions
(i.e. by reproducing the observed value of $M_Z$).
We note that, in  models with CP violation, the usual 
sign ambiguity of $\mu$ becomes a phase ambiguity.

The renormalization group 
running from the GUT scale introduces considerable mass 
splittings in the spectrum of superpartners. 
For example, the gluino becomes much heavier than (roughly triple)
the rest of the gauginos. 
The RG running can also make the scalar quarks quite  heavy, 
\be 
m_{\rm sq}^2(M_Z) \simeq m_0^2 + 6 m_{1/2}^2 + O(M_Z^2),
\ee
especially if $m_{1/2}$ is large. In our 
analysis, the masses of the $A$ and $H$ Higgs bosons are 
functions of the above input parameters and are of particular importance. 
In the CMSSM with large $\tan\beta$, 
they tend to be as heavy
as the sleptons. 
As a consequence, the parameter $\xi$ determining 
the relative importance of the four-fermion operators (\ref{xi}) 
cannot be varied 
over a large range once $\tan\beta$ is fixed. 

In contrast, 
$\xi$ becomes a free parameter in
another version of the MSSM, a CMSSM--type model 
with non-universal Higgs masses (NUHM). 
In this model the squarks and sleptons 
are still mass degenerate at the GUT scale, but the Higgs soft masses are not. 
The model differs from the CMSSM in having two additional 
Higgs mass parameters, which can be taken to be
the mass of the pseudoscalar 
$m_A$ and $\mu$  evaluated at the weak scale, so that the parameter set becomes
\begin{equation}
\left\{\tan\beta,~m_0 ,~ m_{1/2},~m_A,~|\mu|,~\vert A_0 \vert,~ \theta_A,~ 
  \theta_\mu\right\}.
\end{equation}
Relaxing the scalar mass  universality 
brings an interesting new degree of freedom 
into weak-scale phenomenology while preserving 
the flavor-degeneracy of the CMSSM. 
For this study it is important that 
$m_{\rm slepton}/m_A$ and $m_{\rm squark}/m_A$
can be varied over a large range, thus presenting an opportunity to study the 
transition to a regime of dominance for the four-fermion operators.

In what follows, we present our numerical results for the EDMs using 
three two-dimensional slices through the parameter space -- planes of 
$m_{1/2}$--$m_0$ (typical of the CMSSM) and  $\mu$--$m_A$ and $m_A$--$m_0$
(as examples of the NUHM). For each EDM observable,
we plot one CMSSM figure with
$\theta_\mu =0$ and $\theta_A = \pi/2$. 
Note that the phase of $A$ runs in the RGEs, so that at 
low energies, the phases of the various $A_i$ are different
and those of the squarks are typically $\ll \pi/2$, whereas the 
phases of $A_{\tilde l}$ remain relatively large.
In contrast, the phase of $\mu$ does not run so that
$\theta_\mu(EW) = \theta_\mu(GUT)$.
We also plot for each EDM,
three different NUHM cases:
one for  $\theta_\mu = 10^{-3} \pi$ and $\theta_A = 0$ and two
different parameter planes for 
$\theta_\mu =0$ and $\theta_A = \pi/2$. 
In all cases we fix $\tan\beta= 50$ and $|A_0|=300$ GeV
as well as $m_t =175 $~GeV for the pole mass, and 
$m_b = 4.25 $~GeV for the
running $\overline {MS}$ mass evaluated at $m_b$ itself.
Note that our results for the EDMs are not particularly sensitive
to the choice of $|A_0|$.

In all cases, we apply a set of phenomenological constraints that we describe 
briefly. We apply the constraints on new particles from direct LEP searches, namely
$m_{\chi^\pm} > 104$~GeV~\cite{LEPsusy} and Higgs mass 
limits~\cite{LEPref} of $m_h > 114$~GeV. 
The light Higgs mass  has been computed using FeynHiggs \cite{fh}.
We also
require the calculated branching ratio for $b \rightarrow s \gamma$ 
to be consistent
with experimental measurements~\cite{bsg}.
For each set of parameters, we calculate the relic density 
of neutralinos having set the CP-violating phases to 0.
As such the domains we display for a given relic density are
only indicitive.  However,  we note that in the CMSSM (and more generally
when sfermion masses are not too degenerate at the weak scale), the relic density 
of binos is not significantly affected by these phases \cite{WWII}.
For the relic density of neutralinos $\chi$, we show regions of the parameter
planes with $0.1 \le \Omega_\chi h^2 \le 0.3$.  We use this more
conservative range in addition to that suggested by the recent WMAP
data~\cite{WMAP} ($0.094 \le \Omega_\chi h^2 \le 0.129$) for clarity
in the figures.  The effect of the WMAP densities in the CMSSM was discussed
in \cite{cmssmmap} and in the NUHM in \cite{efloso}. We also require that the lightest
supersymmetric particle (LSP) be a neutralino.
For further details on these constraints
see \cite{eoss4}.

%All three sets of figures have shaded areas that
%correspond to: an acceptable neutralino dark matter region, a WMAP-preferred 
%region, a $b\rightarrow s\gamma$ excluded region, and a 
%stau-LSP excluded domain\footnote{Note that these 
%domains are indicative only, as
%the corresponding analysis ignores the effects of phases.}.
%A color-coded (grey-scale) legend for these 
%regions is provided in Fig.~4, in the order described above. 
%In addition, the first three Tl EDM plots have two bands that indicate 
%the cancellation regions where: $d_{\rm Tl}\simeq 0$ in blue (dark grey);
%and $d_e\simeq 0$ in green (lighter grey).

\subsection{Thallium EDM}

\begin{figure}
 \centerline{
   \epsfig{file=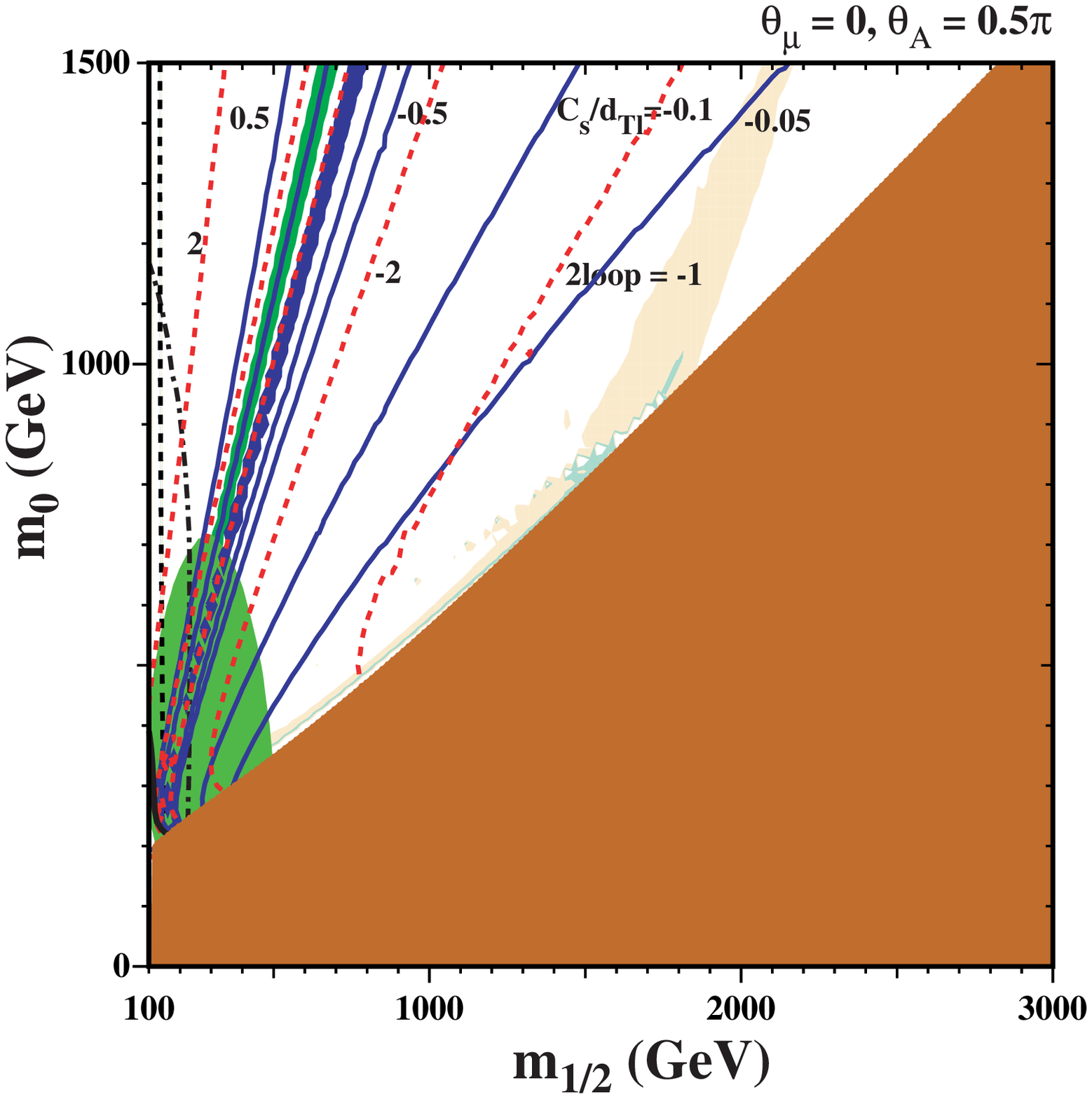,height=8.5cm}
   \epsfig{file=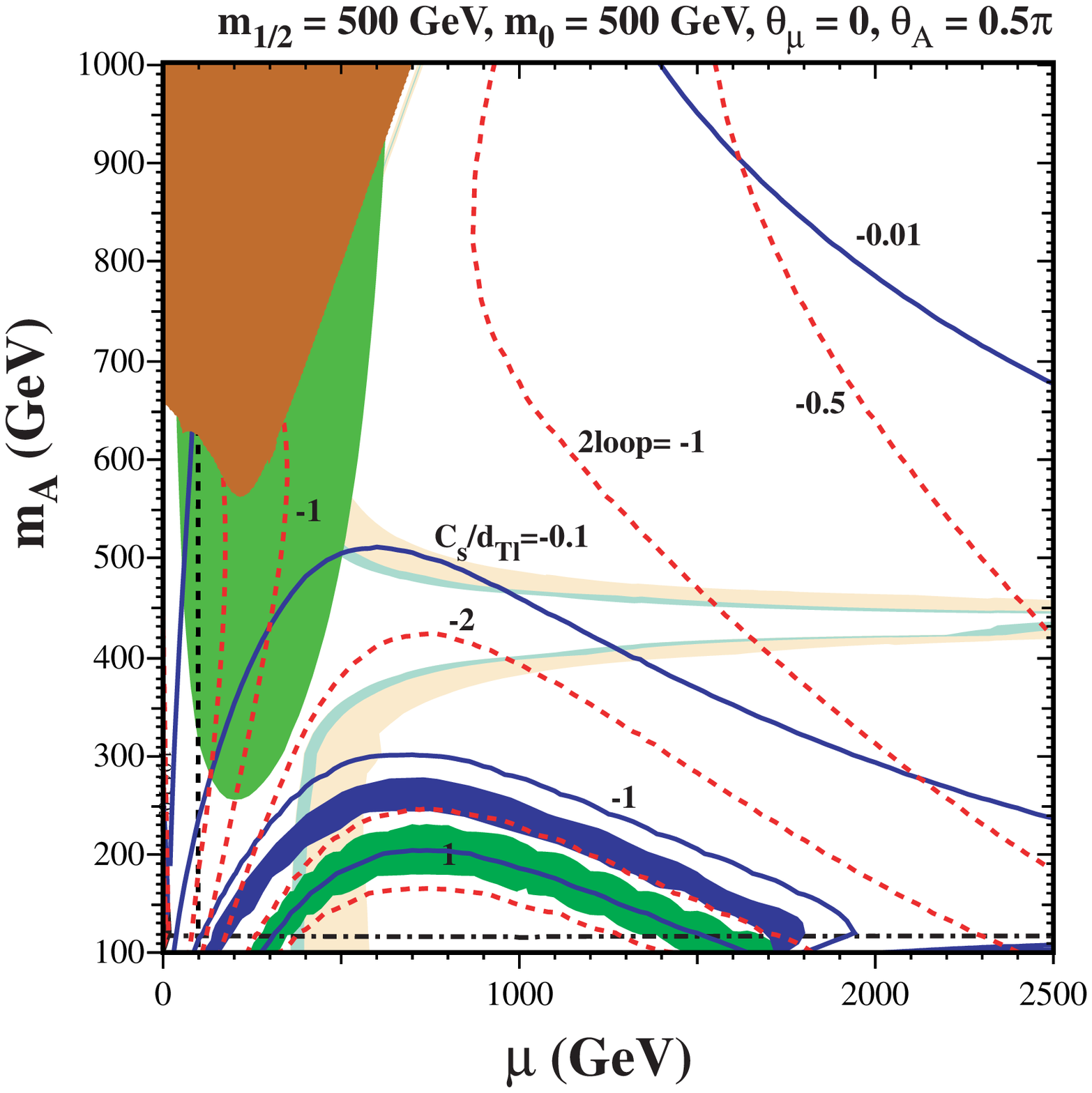,height=8.5cm}
   }
 \vspace*{0.2cm}
 \centerline{
   \epsfig{file=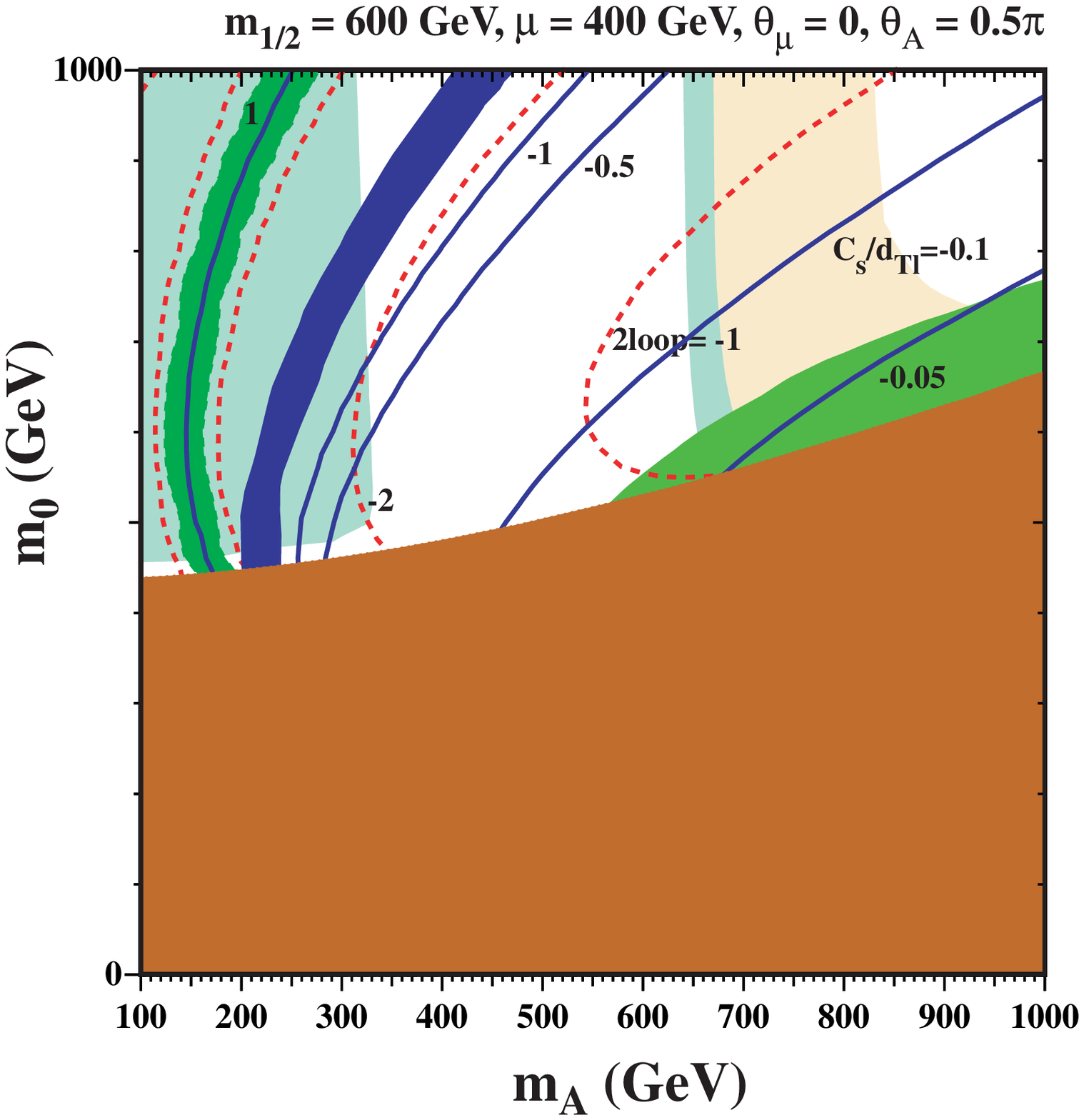,height=8.5cm}
   \epsfig{file=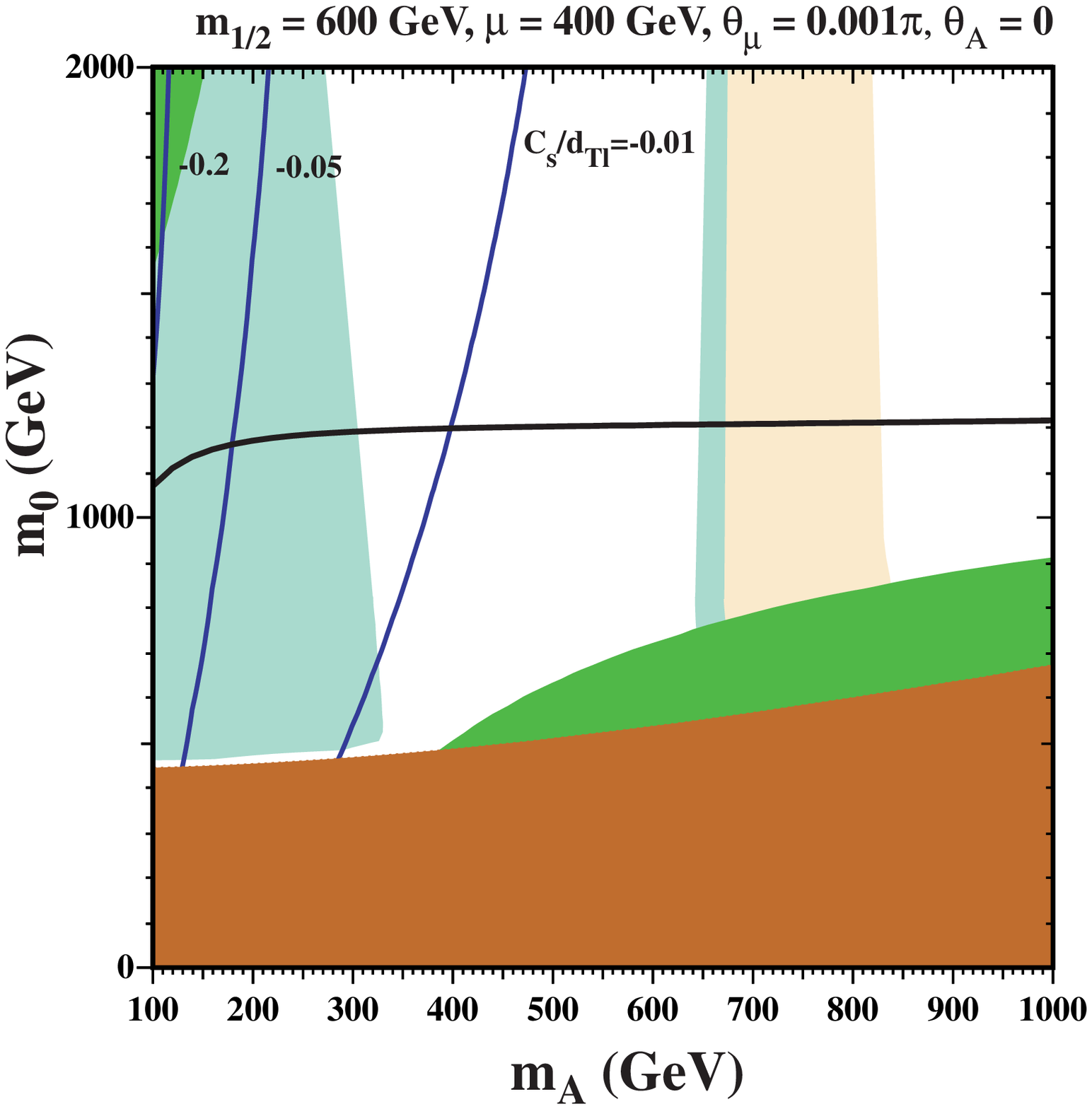,height=8.5cm}
   }
 \vspace*{0.2cm}
 \centerline{
   \epsfig{file=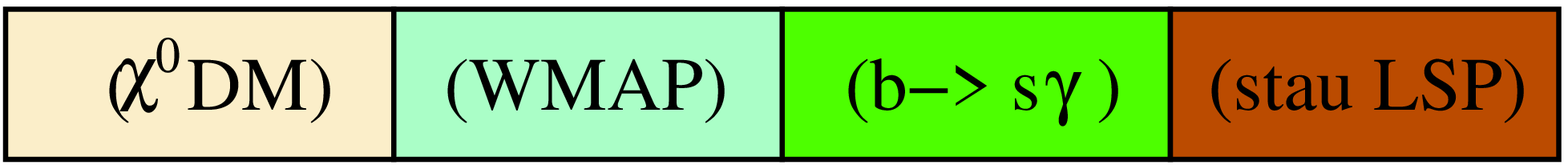,height=0.8cm}
   }
 \newcaption{\small Contributions of four-fermion operators ($C_S$) and the 
electron EDM ($d_e$) to the EDM of thallium as functions of 
the SUSY parameters. Blue (solid) lines are 
contours of constant $d_{\rm Tl}(C_S)/d_{\rm Tl}$, and red (dashed)
lines are the contours of constant $d_e^{2-loop}/d_e^{total}$. A thick solid 
black line marks the current experimental constraint on $|d_{\rm Tl}|$, 
and black dashed and dot-dashed lines represent constraints from 
direct searches for charginos and the Higgs. The legend indicates the
color(grey-scale)-coding of the excluded regions and preferred domains
as described in the text. }
\end{figure}

As emphasized in section 4.1, the thallium EDM receives two main
contributions at large $\tan\beta$, from $d_e$ and $C_S$, while the 
EDM of the electron itself can be generated in different ways at both
one- and two-loop order. Figure 4a plots contours
of constant $d_{\rm Tl}(C_S)/d_{\rm Tl}$ (solid, blue)
and $d_e^{2-loop}/d_e^{total}$  (dashed, red) in
the CMSSM as a function of $m_{1/2}$ and $m_0$. 
The GUT-scale phase of $A_0$ is set to be maximal, 
$\theta_A = \pi/2$, while $\theta_\mu = 0 $. 
The dark (red) shaded lower-right half of the plane corresponds 
to a stau-LSP region which is excluded if the LSP is stable.
The medium (green) shaded region is excluded by
$b \to s \gamma$ (computed with non-zero phases). In the broad light
shaded region the relic density of neutralinos is 
$0.1 < \Omega h^2 < 0.3$ (computed in the absence of phases)
and the thin slightly darker region shows the WMAP relic density.
In this figure, the relic density is determined primarily by
co-annihilations with the nearly degenerate ${\tilde \tau}$
and/or strong s-channel annihilation through the heavy Higgses $H,A$.
In almost all of the unshaded portions of this figure, the relic density is
too high and these parameter choices would be excluded if the LSP is stable.
Also shown by the nearly vertical black dashed line is the LEP bound on the 
chargino mass (smaller values of $m_{1/2}$ are excluded) and the 
black dot-dashed curve is the LEP bound on the Higgs mass (which 
also exlcudes small values of $m_{1/2}$ and $m_0$).

The importance of $C_S$ and $d_e^{2-loop}$ grow as one
moves to lower $m_{1/2}$ and higher $m_0$ 
as indicated by the blue (solid) and red (dashed) lines. In particular,
the contribution from $C_S$ changes dramatically, from merely a  
5\% effect to a 100\% contribution, while the two--loop contributions
are generically large. Moreover, since the 
effects of $C_S$ and $d_e^{2-loop}$ have the opposite sign relative
to the one-loop contribution to $d_e$, the overall consequence is
a significant {\em reduction} of the predicted observable thallium EDM 
over a large portion of the parameter space. In fact, the total 
EDM of thallium is generically several times smaller 
than either $d_{\rm Tl}(C_S)$ or $d_{\rm Tl}(d_e^{total})$ individually. 
The band in parameter space where there is greater than 80\% cancellation 
between $C_S$ and $d_e$ is shaded in blue (dark grey), i.e., 
$d_{\rm Tl}(C_S)/d_{\rm Tl} >5$.
In the center of this band, $d_{\rm Tl}$ actually vanishes.
Our results for $C_S$ are weakly dependent on $\kappa$.
The 50\% uncertainty in $\kappa$, translates into a 10\%
uncertainty in $C_S$. This lack of sensitivity to $\kappa$ can be inferred from
Eq.~(\ref{cstl}) where the first term in the square bracket is dominant.

As noted  earlier, the importance of the two-loop $d_e$ contribution
stems primarily from stau loops since the staus 
are generically lighter than the rest of the 
squarks and sleptons. The left-most band, shaded green (light grey),
in the parameter space shows the region where 
these two--loop contributions nearly screen the one--loop
effects in $d_e$ (to 95\%). In this band, $d_e^{2-loop}/d_e > 20$.
In the center of the band, $d_e = 0$ and the observable atomic 
EDM is given entirely by $C_S$. Note that $C_S$ is dominated by
vertex corrections, $C_{ij}^{({\rm vc})}$, and the contribution due to
$H$-$A$ mixing is not very important.

Turning now to the NUHM, we observe that the cosmologically 
preferred region of parameter space can be substantially wider. 
Figure 4b presents the thallium EDM results  
in the $\mu-m_A$ plane, and one can see that even for 
$m_A$ as high as 500 GeV $C_S$ may contribute 10\% of the total 
thallium EDM, while  for  $m_A< 300$ GeV $C_S $ often provides 
the dominant contribution. Here, we have fixed
$m_{1/2} = m_0 =500$~GeV. For this choice of parameters,
positivity of the Higgs soft masses at the GUT scale would require
$\mu \la 1600$~GeV \cite{nuhm2}.  In this figure, the ${\tilde \tau}$-LSP region
is now in the upper left corner and the region excluded by $b \to s \gamma$ 
extends down from that. The region with the relic density between 0.1 and 0.3
are the two funnel-like strips which become nearly horizontal at large $\mu$. 
This region is allowed by the rapid s-channel annihilation of 
$\chi$'s through heavy $H,A$. Note that $m_{\chi} \approx
 0.4 m_{1/2} \approx m_A/2$ here.  Between these
two strips, the relic density is less than 0.1 and so is not excluded.
The WMAP region corresponds to the thin strips in the interior of the 
funnel.  Once again, we see the blue band with $d_{\rm Tl}(C_S)/d_{\rm Tl} >5$
at $m_A \la 250$~GeV which now intersects the WMAP region
at $\mu \sim 400$~GeV and $m_A \sim 200$~GeV.  At slightly lower
$m_A$, the 2-loop contribution cancels the 1-loop contribution to $d_e$
and $d_{\rm Tl} = d_{\rm Tl}(C_S)$. 

 Perhaps the most interesting slice 
to consider is the $m_A-m_0$ plane plotted in Fig.~4c. 
{}From Fig. 4b, we see that in order to maximize the size of the WMAP cosmological
region, we should choose $\mu = 400$~GeV. Here, $m_{1/2}$ is fixed to
600 GeV. Indeed, in Figs 4c and 4d, we see a large region with 
$m_A \la 300$~GeV (and $m_A \sim 650$~GeV) where $\Omega_\chi h^2 \approx 0.11$. 
For intermediate values of $m_A$ the relic density is small and a dark matter
candidate is not provided. The ${\tilde \tau}$-LSP region now excludes
the lower portion of the figure and $b \to s \gamma$ excludes a small area
at higher $m_0$ when $m_A$ is large.

In Fig. 4c, one 
can see that over most of the cosmologically preferred region, 
$C_S$ provides the dominant 
contribution to the atomic EDM. 
It follows that nowhere within this region can the observable 
EDM be interpreted purely in terms of $d_e$ alone. Moreover, one
observes that within the same cosmologically preferred region 
there is a large band of the parameter space where the two contributions
to the total atomic EDM cancel, implying no constraint on the CP-violating phase. 
Once again, 50\% changes in $\kappa$ lead to 10\% changes in $C_S$.
The final figure (4d) presents the same slice of parameter space
within the NUHM as Fig.~4c for $\theta_A=0$ and $\theta_\mu = 10^{-3} \pi$. 
In this regime, the one-loop contribution to $d_e$ is $\tan\beta$--enhanced,
and provides the dominant effect. We see that even for $\th_\mu \sim 
{\cal O}(10^{-3})$ the domain of parameter space with $m_0 \la 1100$ GeV 
is excluded by the thallium EDM bound as shown by the thick solid black curve. 
The contribution of $C_S$
reaches 20\% on this plot for low $m_A$ and high values of $m_0$. 
Clearly, in this case, over the range of parameter space that is 
potentially accessible at future colliders, the one-loop electron EDM
provides the leading contribution to the atomic EDM of thallium.  
Here, we note that when $\theta_\mu \ne 0$, the sensitivity to 
$\kappa$ is larger and a 50\% change in $\kappa$ leads to a comparable
change in $C_S$.

In summary, this numerical analysis indicates that although 
$C_S$ plays a very important role for paramagnetic EDMs at large $\tan\beta$, 
the precise values are somewhat (a factor of a few) smaller than one 
would naively expect from the simple scaling arguments presented 
in Sec.~2. The main reason for this is that $C_S$ is 
driven mostly by squark loops while $d_e$ also receives contributions 
from selectrons, and as a consequence $C_S$ tends to be further 
suppressed due to the generically heavy masses of the squarks. Another 
important factor is the renormalization group evolution 
from the GUT scale down to the electroweak scale that considerably
suppresses the size of Im$A_t$ relative to Im$A_e$ leading to an 
additional suppression of $C_S/d_e$.

\subsection{Neutron EDM}

\begin{figure}
 \centerline{
   \epsfig{file=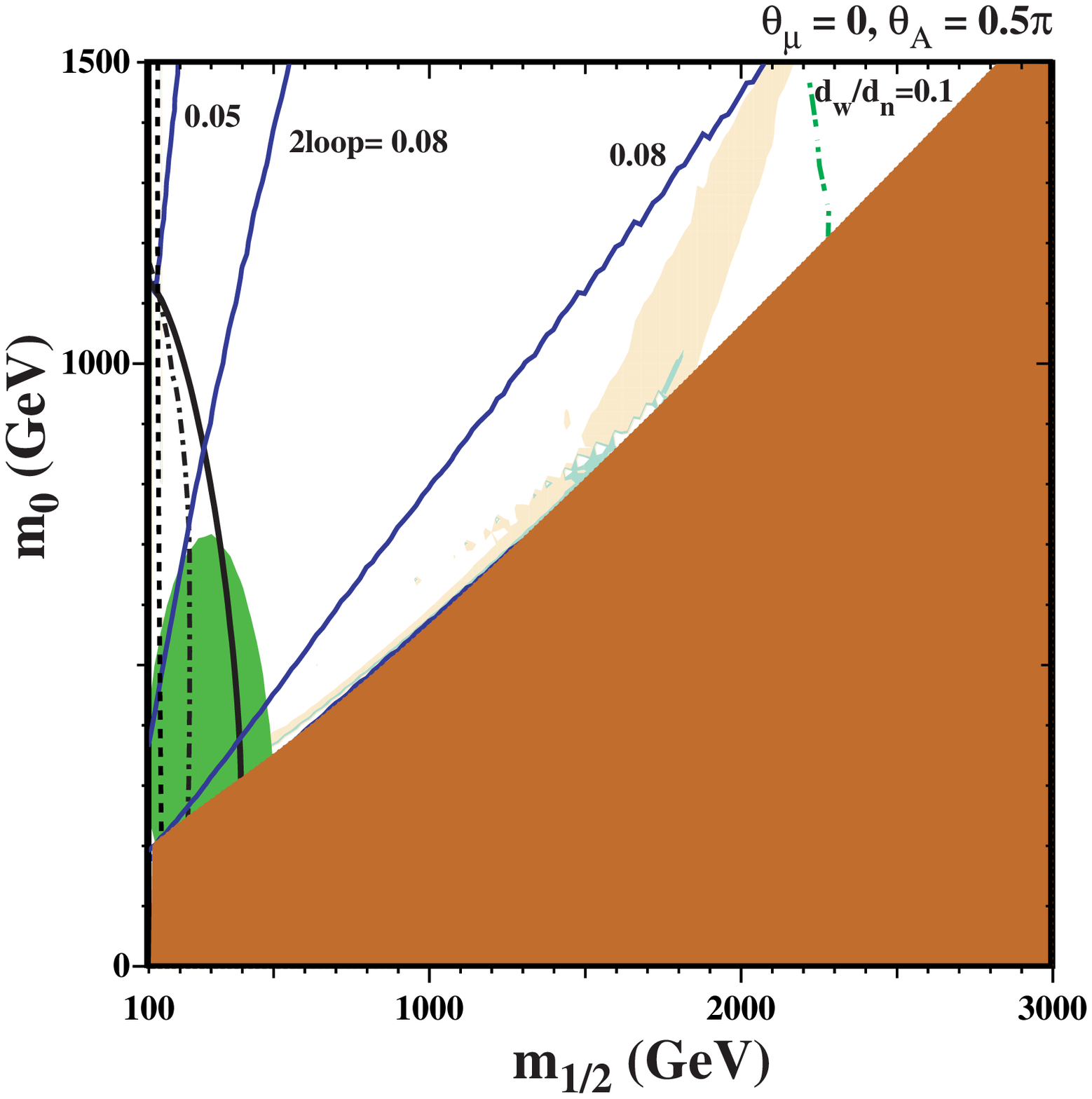,height=8.5cm}
   \epsfig{file=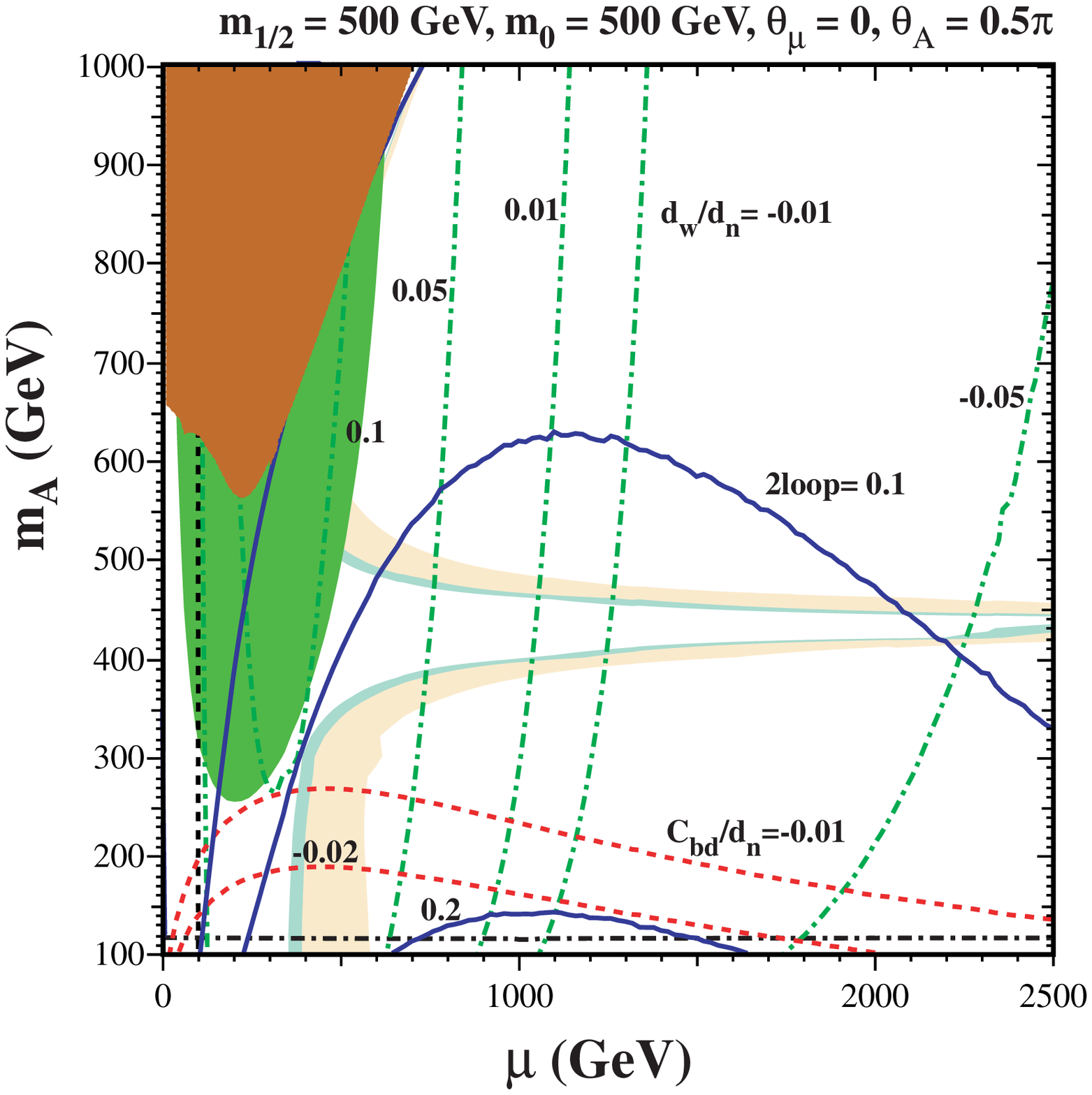,height=8.5cm}
   }
 \vspace*{0.2cm}
 \centerline{
   \epsfig{file=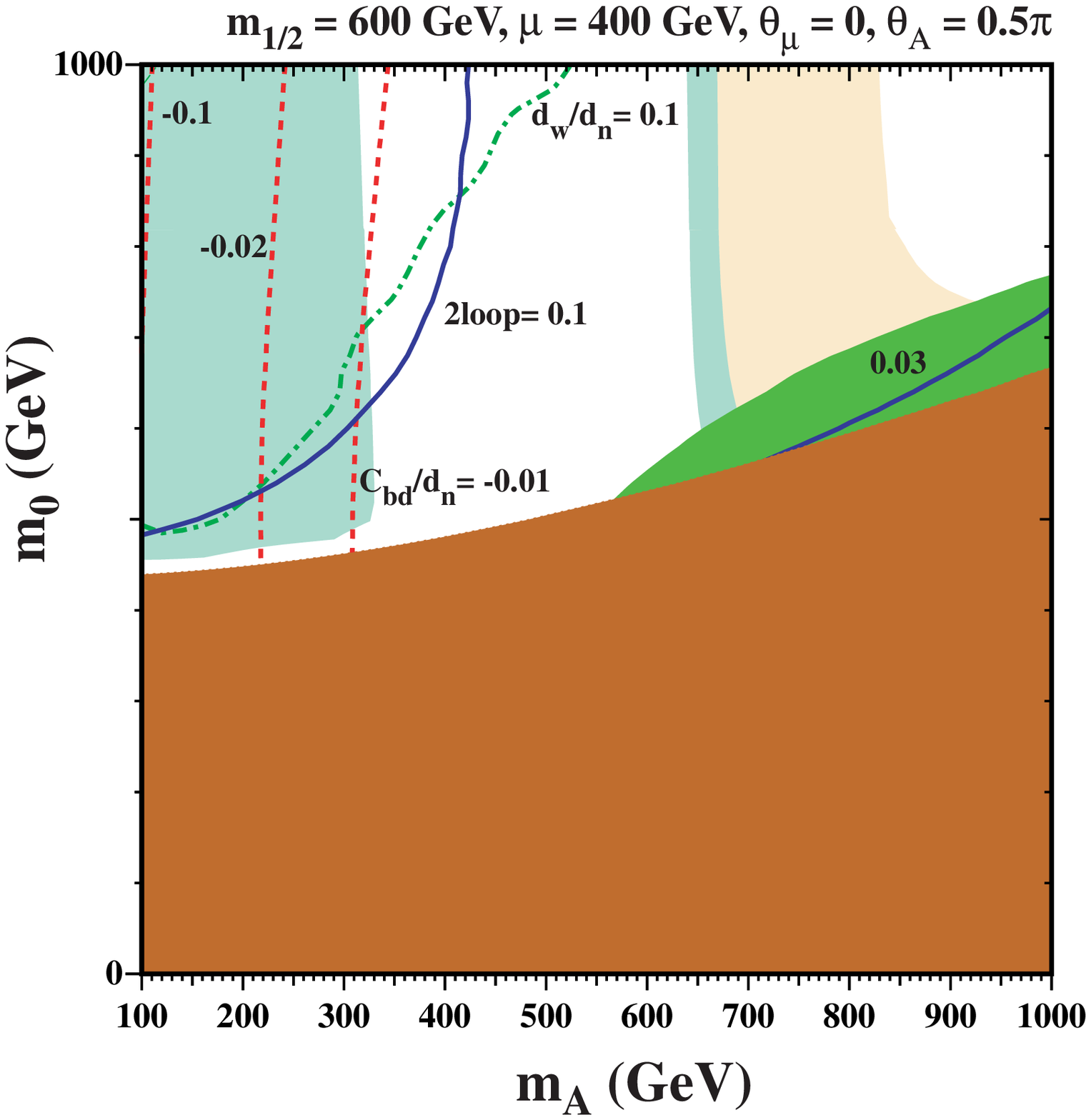,height=8.5cm}
   \epsfig{file=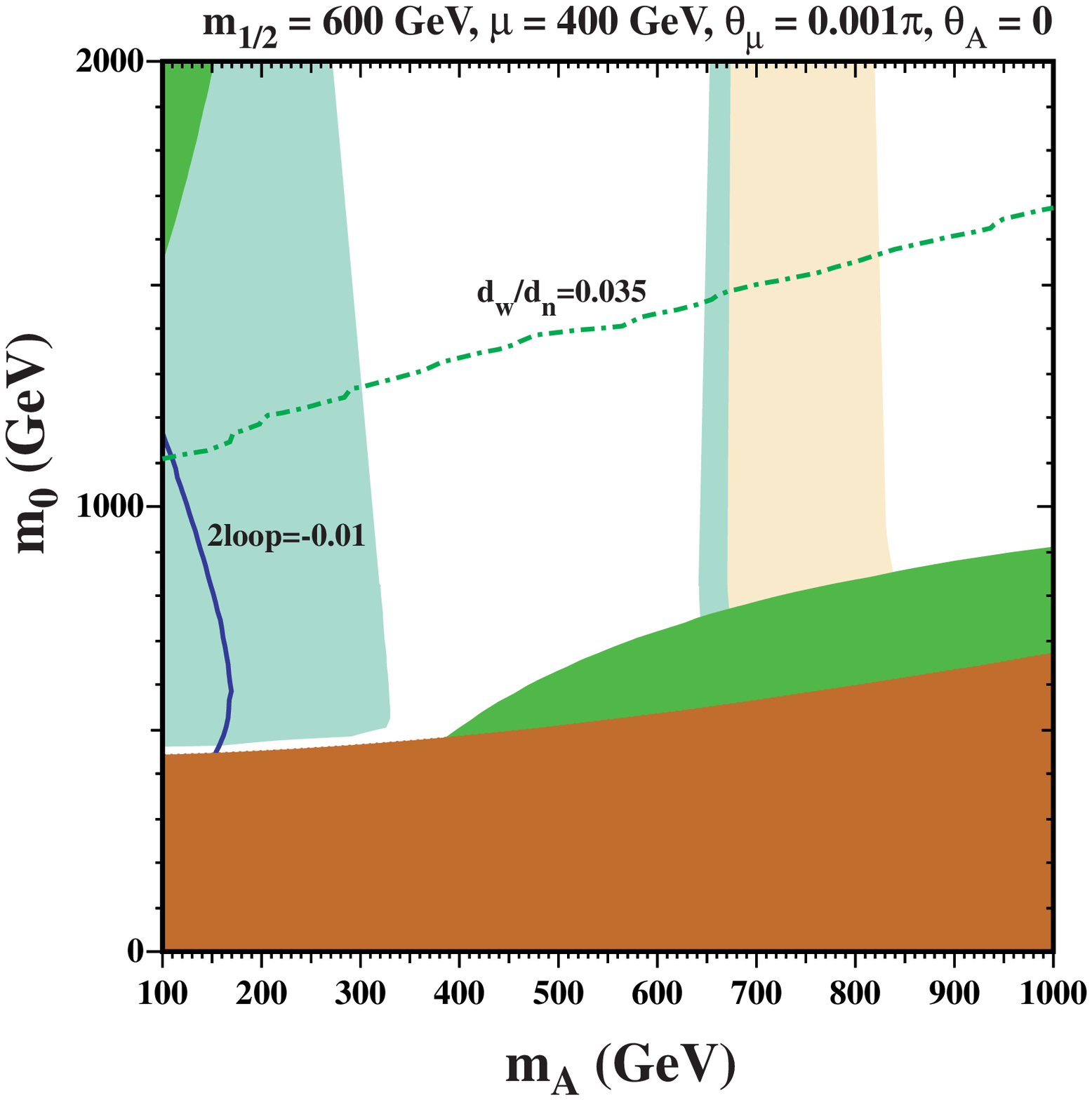,height=8.5cm}
   }
 \newcaption{\small Contributions of the Weinberg operator, 
$d_n(w)\equiv d_w$, and four-fermion operators, $d_n(C_{ij})$ denoted as 
$C_{bd}$ here, to the neutron EDM as functions of SUSY parameters. 
Blue (solid grey) lines denote contours of constant 
$d_{n}(d_i^{2-loop})/d_{n}$, while red (dashed)
lines are contours of constant $d_n(C_{bd})/d_n$, and finally
green (dot-dashed) lines are contours of constant $d_w/d_n$. 
The thick solid black line delineates the experimental bound on
$|d_{n}|$, and black dashed and dot-dashed lines represent constraints from 
direct searches for neutralinos and the Higgs.}
\end{figure}

Repeating the same sequence of plots for the neutron EDM, we observe important 
differences with respect to thallium. In all four panels of Fig. 5, the phenomenological
and cosmological considerations are identical to those described for 
Fig. 4. In the CMSSM, with 
$\theta_A = \pi/2$, the current experimental bound, shown as the thick black
curve, for $d_n$ excludes much more of the plane
than that for  $d_{\rm Tl}$ (due to the absence of any significant cancellations). 
Indeed, for low values of $m_{1/2}$ around 100 GeV, scales for 
$m_0$ above 1 TeV are probed by the experimental constraint. 
In the other planes, the choices for $m_{1/2}$ and $m_0$ and/or $\mu$
were made so as to satisfy ab initio  the current experimenatal bounds over the 
displayed plane. 
Furthermore, 
there is little sensitivity to $\th_\mu=10^{-3}\pi$  compared to that of
the thallium EDM. The contribution of the two-loop quark EDMs and CEDMs
is in general very small, not exceeding 10\% of the total $d_n$. 
Moreover, of particular relevance here, a scan over the 
parameter space of the CMSSM and NUHM reveals that 
the four-quark interactions and the Weinberg operator 
contribute at the 10\% level at most.   
Both the Weinberg operator and the two loop contributions $d_i^{2-loop}$ 
scale as (two-loop factor)$\times\tan\beta$ and therefore it is natural that 
they both contribute at a similar level. 
The contribution of four-fermion operators is dominated by $C_{bd}$ and 
increases 
when $m_0$ is increased and/or $m_A$ is lowered. Although outside the
parameter regime included in the plots, it is worth noting that
a factor of 3 increase in the maximum value of $m_0$ to around 3 TeV 
in Fig.~5c (at low $m_A$) would bring the 
$C_{bd}$ contribution to a level comparable with $d_n(d_i,\tilde d_i)$. 

We conclude that for all $m_0 < $ 1 TeV the neutron EDM at large $\tan\beta$
can be reliably calculated in terms of the 
the EDMs and CEDMS of quarks. The hadronic 
uncertainties associated with the matrix elements of four-fermion
and Weinberg operators, which we can only estimate within a factor of 3, 
do not significantly disrupt the predictions for  $d_n$. 
This means that even at large $\tan\beta$ the calculation of $d_n$ can
provide meaningful constraints on specific combinations of the quark
EDMs and CEDMs.

\subsection{Mercury EDM}

\begin{figure}
 \centerline{
   \epsfig{file=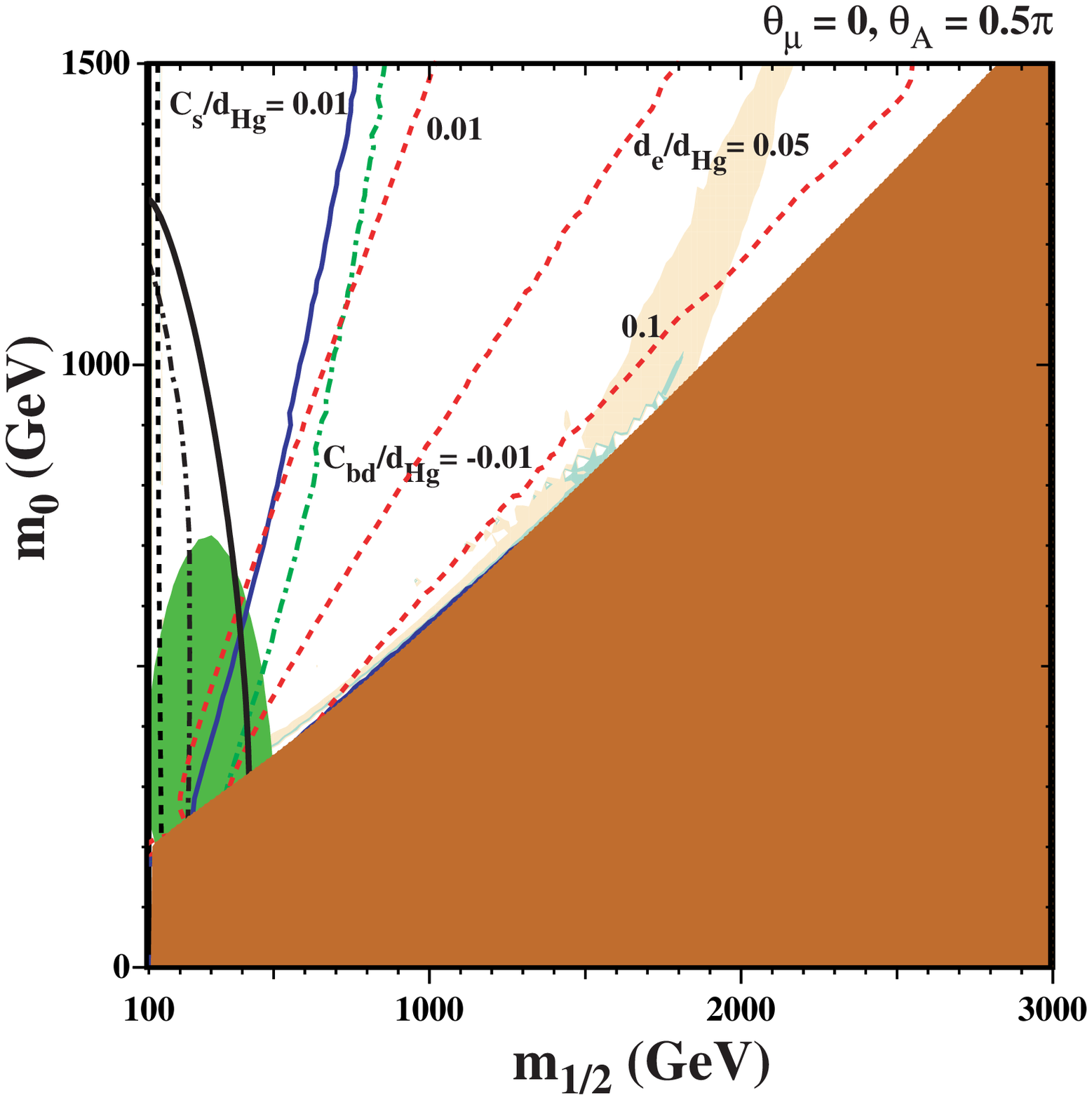,height=8.5cm}
   \epsfig{file=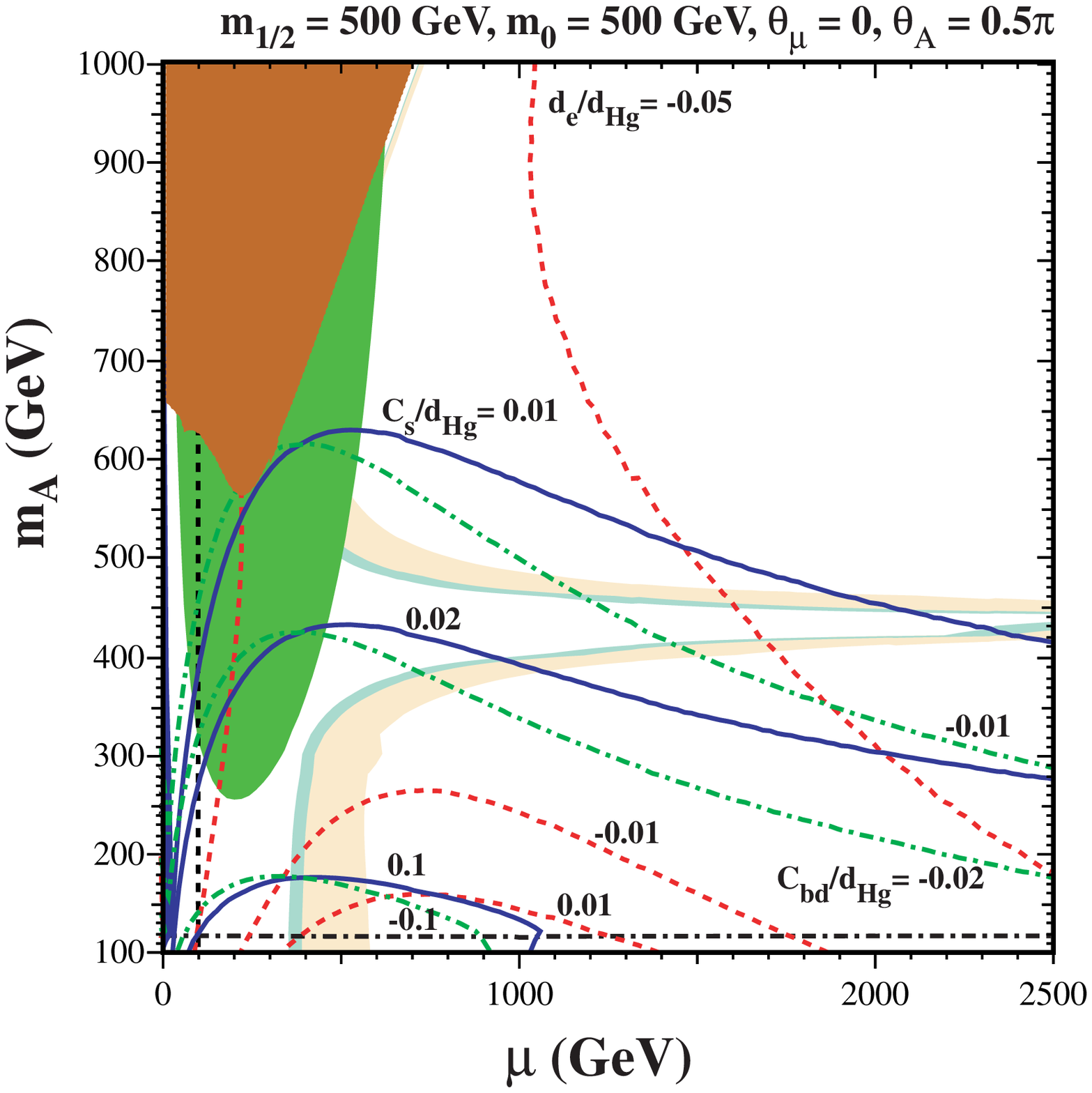,height=8.5cm}
   }
 \vspace*{0.2cm}
 \centerline{
   \epsfig{file=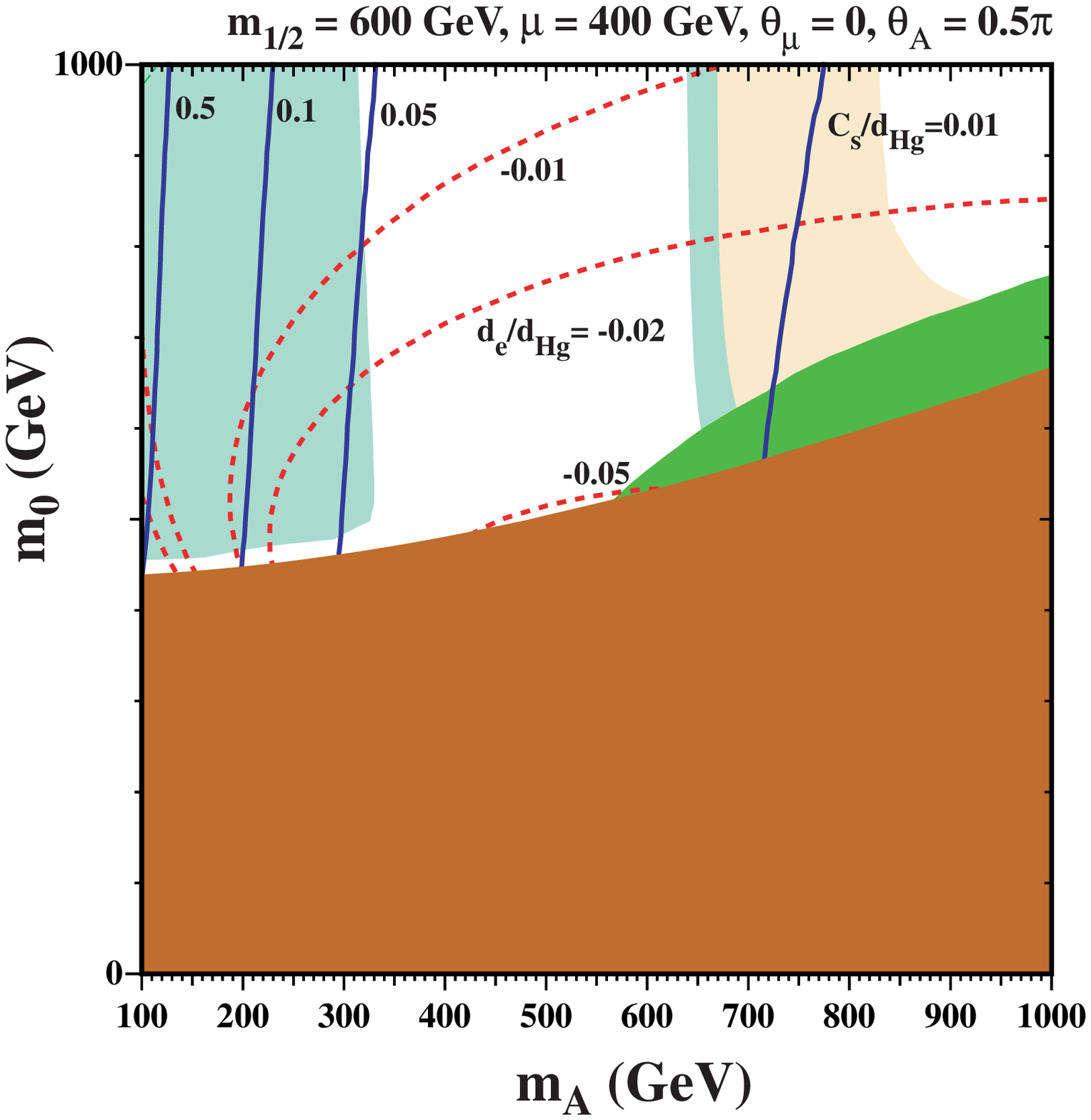,height=8.5cm}
   \epsfig{file=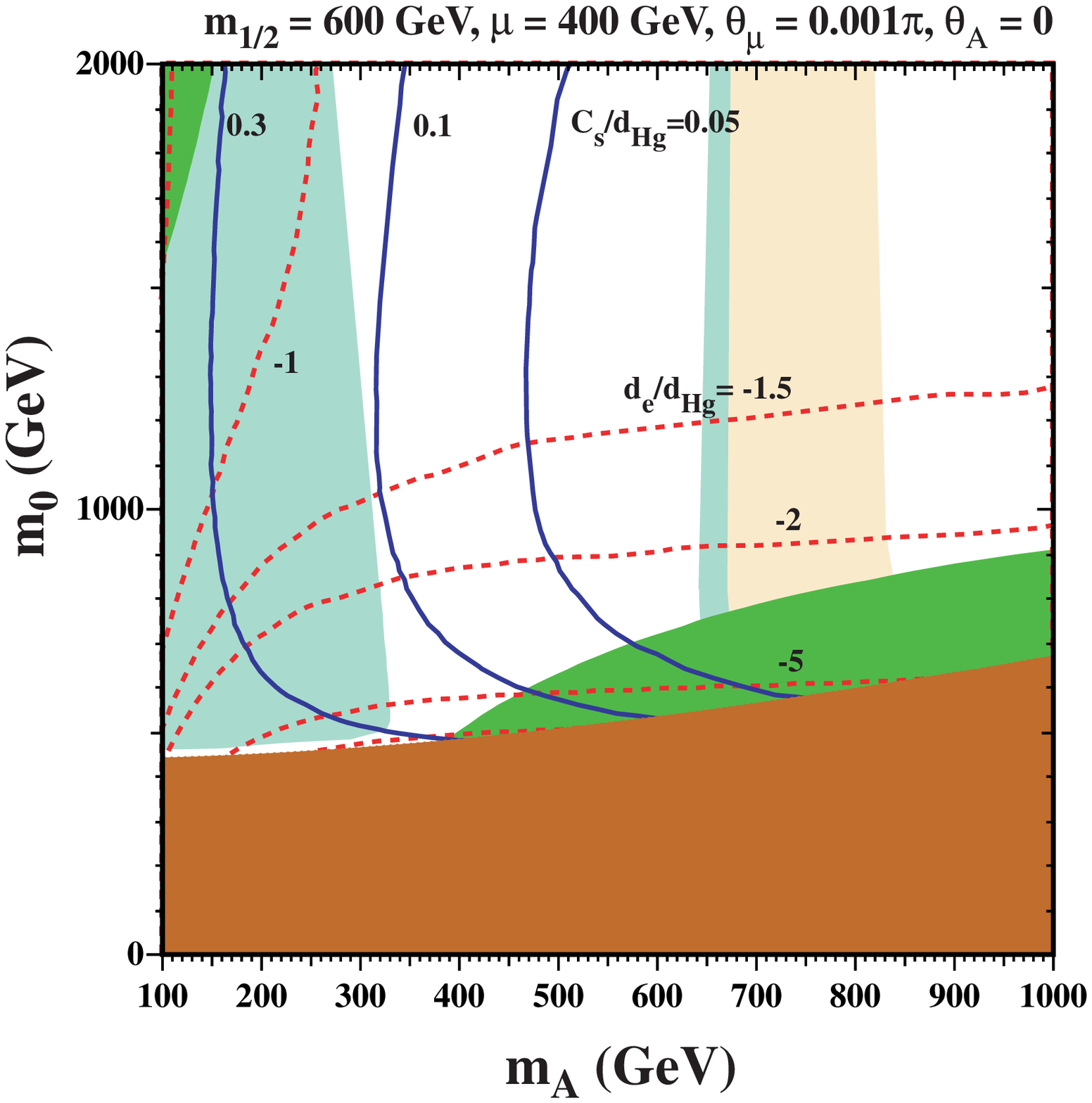,height=8.5cm}
   }
 \newcaption{\small Contributions of the four-fermion 
operators and the electron EDM to the EDM of mercury as 
functions of the SUSY parameters. Blue (solid grey) lines denote
contours of constant $d_{\rm Hg}(C_S)/d_{\rm Hg}$, while red (dashed)
lines are the contours of constant $d_{\rm Hg}(d_e)/d_{\rm Hg}$ 
(indicated as $d_e/d_{\rm Hg}$), and green (dot-dashed) lines  are  
contours of constant $d_{\rm Hg}(C_{q_1q_2})/d_{\rm Hg}$ 
(indicated as $C_{bd}/d_{\rm Hg}$ on the plot). 
The thick solid black line denotes the 
experimental bound on $|d_{\rm Hg}|$, while 
black dashed and dot-dashed lines represent constraints from 
direct searches for neutralinos and the Higgs.}
\end{figure}

In many ways, the mercury EDM occupies an intermediate position between 
$d_n$ and $d_{\rm Tl}$. 
Results for the Hg EDM are shown in Fig. 6.
Once again, the phenomenological
and cosmological considerations are identical to those described for 
Fig. 4.
In the CMSSM with $\th_A \neq 0$ and 
$\th_\mu = 0$, the current experimental constraint on
$d_{\rm Hg}$ shows even greater sensitivity to $m_0$
than does $d_n$. On the other hand, the contribution of 
four-fermion operators to $d_{\rm Hg}$ is relatively less important 
than for $d_{Tl}$, but more important than for $d_n$. 
It is interesting to note that the contribution of $C_S$ 
has a similar magnitude, but the opposite sign, to that of
$C_{q_1q_2}$, and over a large part of the parameter space 
these two effects tend to compensate each other. An exact cancellation 
would be an accident, as $C_{q_1q_2}$ has considerable hadronic uncertainties 
associated with the calculation of the 
pion-nucleon CP-violating coupling and the Schiff 
moment. The contribution from $C_P$ is negligible.

The effects of four-fermion operators become much more pronounced in the 
NUHM model. Fig.6c shows that for low $m_A$ 
the mercury EDM receives contributions of up to 50\% 
from $C_S$ and $C_{q_1q_2}$ (for the chosen parameters 
$d_{\rm Hg}(C_S) \approx - d_{\rm Hg}(C_{q_1q_2})$ and only $C_S$ is 
shown on the plot). 
These contributions drop quickly with increasing 
$m_A$ and for $m_A>$ 250 GeV do not exceed 10\% of the total EDM.
It is also important to note that for both  the 
CMSSM and the NUHM the electron EDM remains subdominant for 
$\th_A \neq 0$ and $\th_\mu = 0$. This provides us with some 
confidence in the calculation of the mercury EDM at $\th_A \neq 0,~\th _\mu = 0$
in the range $m_A > 250$ GeV and $m_0<$ 1.5 TeV where 
{\em both} $d_e$ {\em and} four-fermion operators provide contributions
of less than 10\%. We conclude that in this part of the parameter space  
the mercury EDM is dominated by contributions from
$(\tilde d_u - \tilde d_d)$.

This situation changes when 
we consider the case of $\th_A = 0,~\th_\mu \neq 0$,
shown in Fig. 6d. One can readily see that the electron EDM becomes 
important everywhere within the domain of parameter space included in 
the plot, and its inclusion leads to a significant {\em reduction} in 
the total mercury EDM. In practice, however, in this regime $d_{\rm Tl}$
is dominated by $d_e$ and, on taking the bound on $d_{\rm Tl}$  into 
account, the effect of $d_e$ on $d_{\rm Hg}$ will be suppressed. 
We note finally that, unlike the previous 
observables, at $\th_A = 0,~\th _\mu \neq 0$ the mercury EDM 
also depends sensitively on $C_S$ and the four-fermion contributions 
to the Schiff moment.

\section{Conclusions}

In this paper, we have presented the results of SUSY-EDM calculations 
in the large $\tan\beta$ regime. Large values of 
$\tan\beta$ are motivated by a 
variety of theoretical models and the SUSY Higgs 
searches at LEP also suggest that $\tan\beta\ga 5$. Our main goal was to combine 
the existing one and two-loop 
calculations of the EDMs of the constituents (electrons, quarks, etc.) with 
the effects of four-fermion CP-odd interactions, the importance of which was
recently emphasized  in Ref. \cite{LP}. We concentrated
on comparing the relative size of these contributions, while
a more detailed analysis of the constraints on SUSY phases 
will appear elsewhere \cite{DLOPR2}.

We have compiled the most accurate formulae to date
 to obtain the three observables:
the EDMs of the neutron, mercury and thallium as functions 
of the Wilson coefficients, 
which in turn are expressed in terms of SUSY parameters.
Mostly these are compilations of existing 
QCD/nuclear/atomic calculations, although in certain cases 
such as $d_n(C_{bd})$ and $d_{\rm Hg}(C_{bd})$, 
new QCD estimates have been  obtained. 
On the SUSY side, we have performed 
a complete calculation of 1--loop corrections to the 
quark and lepton masses in the large $\tan\beta$ regime,
including the effects of SUSY CP-violating phases.
These mass corrections (or rather their imaginary parts) 
determine the strength of the  
CP-violating Higgs-fermion interaction, the leading source 
for the CP-odd quark-quark and quark-electron interactions. At the 
subleading level, 
only the effects of scalar-pseudoscalar mixing were taken into account 
as these are important 
for low pseudoscalar masses $m_A$, although a large class of box diagrams 
can contribute as well. We also included the 
stau-loop contribution to the two-loop corrections to the EDMs, 
and recalculated the corresponding sign to firmly establish the pattern of 
interference between one- and two-loops.

The impact of the four-fermion operators is most striking for the
thallium EDM. We observe that for 
low $m_A$ and high $m_0$ the electron EDM is no longer dominant, 
and $C_S$ provides as 
large a contribution (or possibly even larger) to $d_{\rm Tl}$ 
as does $d_e$. 
A similar conclusion will apply
to other paramagnetic species as well. 
This is an important point  because it illustrates that in 
SUSY models with large 
$\tan\beta$,  the EDM of a heavy paramagnetic atom or 
molecule cannot be entirely attributed to  $d_e$. 
Should the next generation of paramagnetic EDM 
experiments \cite{YbF,PbO} detect a 
non-zero signal, only a combination of positive measurements with 
different atoms or molecules could unambiguously separate the two effects. 
It is also noteworthy that the total electron EDM is 
reduced due to destructive interference between the 
one- and two-loop contributions, particularly due to ${\tilde \tau}$-loops.

In the $C_S$-dominated regime for $d_{\rm Tl}$, the main uncertainty in the 
theoretical calculation comes from the uncertainty of the 
strange quark scalar matrix element over the nucleon. This 
translates into a $ 10\%$ uncertainty 
when the EDM is induced 
by $\theta_A$ and $ 30-50\%$ uncertainty for 
the $\theta_\mu$--induced EDM. 
The transition to dominance 
of $C_S$ for $\theta_\mu\neq 0$ occurs at a very large mass 
scale for the superpartners, and in reality 
for $m_0<\,$2 TeV the contribution of $C_S(\theta_\mu)$
is always subdominant. 
Yet, in the case that the EDM is induced by $\theta_A$, we find that 
the four-fermion operators can easily contribute as much as the 
electron EDM\footnote{We note also that
additional $\tan\beta$--enhanced contributions may 
arise through the electron EDM via two-loop renormalization 
group evolution of the gaugino masses. We will address these effects
in more detail elsewhere \cite{DLOPR2}.}. We also 
find that, in certain parts of the parameter space, 
the physical observable $d_{\rm Tl}$  
is considerably  smaller than the individual contributions. 

In the case of the mercury EDM, there are a variety of different 
contributions that have to be taken into account. We find 
that at large $\tan\beta$ the contribution of $C_S$ 
and other four-fermion operators may become important. 
From our analysis it is clear that the main problem 
with the interpretation of the mercury EDM at 
large $\tan\beta$ is that a rather large 
number of contributions appear with many of them having different 
hadronic/nuclear/atomic matrix elements. Therefore, a 
reliable calculation (i.e. better than 100\% uncertainty) 
becomes problematic at large $\tan\beta$. This contrasts with 
low and moderate $\tan\beta$ where the EDM of 
mercury is usually dominated by the color EDMs of the light quarks, and
the hadronic uncertainties are a less harmful overall factor.
Nevertheless, for $\th_\mu =0$, $m_A \ga $ 250 GeV and $m_0 \la$ 1.5 TeV 
the effects of four-fermion operators and the electron EDM are subdominant 
and a meaningful estimate for $d_{\rm Hg}$ is possible.

We determined that the EDM of the neutron is the least susceptible 
observable, with the 
contributions of four fermion operators 
never exceeding 10\% of the total EDM for $m_0 \la 1$~TeV. 
On the other hand, for $\theta_A \neq 0$ 
the contribution of the Weinberg operator can be quite 
substantial. According to 
our estimates for the matrix elements, it is $\la$ 10\% of
the total EDM in much of the parameter space considered
(if an NDA estimate of $d_n(w)$ is employed, 
this contribution is larger). 
Although subleading, we infer
that at large $\tan\beta$ the calculation of the neutron EDM 
can be prone to 
uncertainties related to the matrix element of the Weinberg operator. 
It is also worth noting that the
dominant, $\tan\beta$-enhanced, contribution to the Weinberg 
operator comes from the color EDM of the $b$-quark. 

Finally, we remark that low values of $m_A\la 300$ GeV can be 
preferred by considerations of the cosmological 
neutralino abundance. This is related to the new 
annihilation channels due to Higgs exchange that open up when $m_A$ is not 
excessively large. In that part of the 
parameter space the effects of four-fermion 
operators are very important, and therefore models that include both  
CP violation and neutralino dark matter should incorporate the effects of 
four-fermion interactions into the EDM constraints.

%Our investigation also included some of the 
%two-loop contributions to $d_i$ and $\tilde d_q$. 
%We find that these two-loop contributions are not 
%so important for the EDM of the thallium atom, while 
%they are very significant for $d_n$ and $d_{\rm Hg}$. Therefore, it is highly
%desirable that the full set of these two-loop 
%EDM diagrams enhanced by $\tan\beta$ be calculated, 
%as at the moment only partial results are available. 

\subsection*{Acknowledgments}
We would like to thank Y. Nir and Y. Santoso for helpful conversations.
The work of DD and KAO  was supported in part
by DOE grant DE--FG02--94ER--40823. The work of MP
was supported in part by NSERC of Canada and PPARC of the UK.
AR would like to thank 
the FTPI at the University of Minnesota for
generous hospitality while this work was completed.

\appendix

\section*{Appendices}

\section{Neutron EDM induced by four-fermion operators}

As discussed in Section~4, the direct QCD sum-rule analysis for 
the neutron EDM becomes problematic for sources of dimension six, due
in part to the presence of unknown condensates. For the 
contribution of the Weinberg operator, one can nonetheless
obtain a useful estimate \cite{DPR} by relating $d_n$ to the neutron
anomalous magnetic moment, $\mu_n$, by a chiral rotation of the
nucleon wavefunction. For similar reasons, we will adopt the same approach
here in estimating the contribution of the four fermion sources to
the neutron EDM.

To proceed, it is convenient to first introduce a redefined 
set of Wilson coefficients, so that
\be
 \de{\cal L} = \tilde{C}_{bd}(m_b) m_b \bar{b} b\bar{d}i\gamma_5 d
   + \tilde{C}_{db}(m_b) m_b \bar{d} d \bar{b}i\gamma_5 b, 
\ee
at the $b$-quark mass scale, where $\tilde{C}=C/m_b$. Below this
scale, the $b$-quarks can be integrated out leading to 
dimension-7 quark-gluon operators.
Making use of the standard low energy theorems we obtain
\be
 \de{\cal L} = -\frac{\al_s(\mu) 
 \tilde{C}_{bd}(\mu)}{12\pi} \bar{d}i\gamma_5 d GG
   - \frac{\al_s(\mu)\tilde{C}_{db}(\mu)}{16\pi} \bar{d} d G\tilde{G}.
   \label{source} 
\ee
The evolution of the coefficients to the hadronic scale $\mu\sim\,$1 GeV
requires knowledge of the anomalous dimensions of these
dimension-7 operators. Since (with hindsight), the overall contribution
of these operators is small, we will approximate this by the anomalous
dimension of $\bar{q}q$, recalling that $\al_s GG$ is RG--invariant. Thus,
we run $\tilde{C}_{bd}$ and $\tilde{C}_{db}$ to $\mu\sim\,1\,$GeV, using 
$\gamma_{\al_sqqGG} \sim 4/9$.

To gain some intuition about the dependence of $d_n$
on $\tilde{C}(\mu)$, we first utilize ``naive dimensional analysis'' 
\cite{GM,W}. The
quark-gluon operators are of dimension-7, so the reduced 
coupling is $M^3/(4\pi)^2$ with $M \sim 4\pi f_{\pi}$. The 
corresponding estimate for $d_n$ is given by $e/M$ multiplied by
the reduced coupling of the source, and so we find
\be
 |d_n({\rm NDA})| \sim  e\ 2.1 \times 10^{-3} \ {\rm GeV}^2 \ 
  \left( \tilde{C}_{bd}(\mu) 
   + 0.75\ \tilde{C}_{db}(\mu) \right),
\ee
on using the usual matching condition $\al_s=4\pi$, i.e. at a very low
(apparent) scale. As it turns out, our final estimate 
is quite close to this, assuming we choose the matching scale 
not too far from 1~GeV.

We now turn to a sum-rule analysis. As alluded to above, a direct
calculation appears intractable due
to the presence of unknown condensates. Therefore, we will pursue instead
an indirect method first considered by Bigi and Uraltsev \cite{BU}, and used
recently within QCD sum-rules to estimate $d_n$ induced by the
Weinberg operator $GG\tilde{G}$. One considers the $\gamma_5$-rotation
of the nucleon wavefunction induced by $\de{\cal L}$, leading to
an estimate for $d_n$ in terms of the corresponding rotation of the
anomalous magnetic moment $\mu_n$:
\be
 d_n \ \sim \ \mu_n \ \frac{\langle N| \de {\cal L} | N\rangle}{m_n
\bar{N} i \gamma_5 N}. \label{rotate}
\ee

In the present context, we require the ratio of the leading 
contributions to the structures {\bf 1} and $i\gamma_5$ in the 
mass sum-rule correlator. We have 
\be
 \int d^4 x e^{ip\cdot x} \langle \et_n(0) \bar{\et}_n(x)\rangle = 
 \frac{1}{16\pi^2}\qq \left[p^2 \ln(-p^2){\rm \bf 1} 
 + f_{\de}(p^2) i\gamma_5 \right], \label{cf}
\ee
where $\et_n$ is the neutron current, which for the mass sum-rule we take
at the ``Ioffe point'' \cite{ioffe}, $f_{\de}$ is the contribution 
induced by the CP-odd souce $\de{\cal L}$ that is to be determined, and 
we have retained only the leading order contribution to the structure {\bf 1}.

\begin{figure}
\begin{center}
\epsfig{file=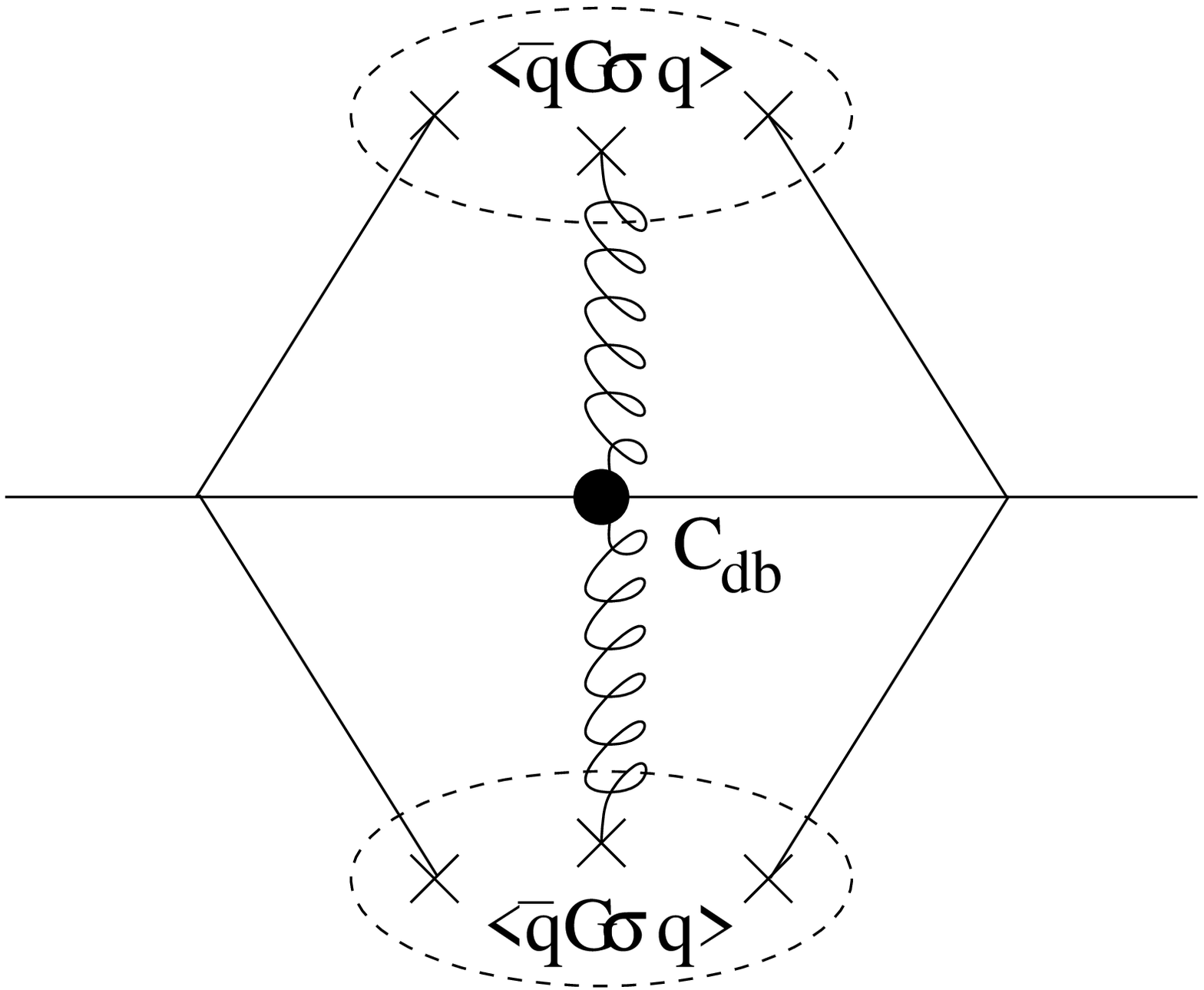, width=5cm, angle=0}
\end{center}  
\vskip .1in
\newcaption{\small\label{fig3}  The leading contribution to the
$\gamma_5$ structure induced by the CP-odd source.}
\end{figure}

Fortunately, in this case the leading contribution 
enters in a tractable form without the need for additional 
perturbative vertices -- i.e. the gluon structure of the 
source $\de{\cal L}$ can be extracted at the nonperturbative level. 
The relevant class of diagrams is exhibited in Fig.~7. Using standard
methods, we calculate the quark propagators in the background of the
two sources in $\de{\cal L}$ making use of the known 
guark-gluon condensate\footnote{We have introduced a tilde atop 
$m_0$ in order to distinguish this QCD condensate from the mass 
of the SUSY scalars.}:
\be
 \langle \bar{q}g_s (G\si) q\rangle = \tilde m_0^2 |\qq | \approx 0.8 {\rm
GeV}^2 \qq.
\ee
One then obtains,
\be
 f_{\de}(p^2) = -\frac{\tilde{m}_0^4}{6p^2} \qq \left(\tilde{C}_{bd} 
 + 0.75 \tilde{C}_{db}\right),
\ee
and we proceed by performing a Borel transform
and comparing the two terms in (\ref{cf}) at the neutron 
mass scale in terms of the Borel parameter $M$. This fixes the ratio of
the two structures in (\ref{cf}) which is the quantity that enters
the estimate (\ref{rotate}), leading to the result,
\be
 |d_n| \sim |\mu_n | \frac{\tilde m_0^4}{6 M^4} \qq \left(\tilde{C}_{bd} 
 + 0.75 \tilde{C}_{db}\right) 
 \sim e\ 2.6 \times 10^{-3} {\rm GeV}^2 \left(\tilde{C}_{bd} 
 + 0.75 \tilde{C}_{db}\right),
\ee
which necessarily comes with a precision estimate not better than
${\cal O}(100\%)$. On re-expresing this result in terms of $C_{bd}$ and 
$C_{db}$, we obtain the result quoted in (\ref{dn3}).

Finally, for completeness, we note that the operators 
$C_{dd}$, $C_{ds}$, $C_{sd}$ and $C_{ss}$ can be generated via
$H-A$ mixing. Their contribution to the neutron EDM can 
be estimated as \cite{KL,PH}:
\be
\frac{d_n(C_{dd})}{C_{dd}}\sim \frac{d_n(C_{sd})}{C_{sd}}\sim 
 2\times 10^{-2}{\rm GeV},\;\;\;
 \frac{d_n(C_{ds})}{C_{ds}}\sim \frac{d_n(C_{ss})}{C_{ss}} 
 \sim 2\times 10^{-3}{\rm GeV}.
\ee

\section{Mercury EDM induced by four-fermion operators}

The main source for the Schiff moment of the mercury nucleus arises through a 
pion-exchange contribution with CP violation in the pion-nucleon vertex, 
\begin{eqnarray}
 {\cal L}_{\pi NN } = \bar g_{\pi NN}^{(0)}\bar N\tau^aN\pi^a
+\bar g_{\pi NN}^{(1)}\bar NN\pi^0+
\bar g_{\pi NN}^{(2)}(\bar N\tau^aN\pi^a - 3 \bar N\tau^3N\pi^0),
\end{eqnarray}
of which the contribution 
of $\bar g_{\pi NN}^{(1)}$ is the most important \cite{P,DS}.
We recall here that the main technical difficulty in the derivation 
of this coupling, as induced 
by color EDM operators, is the existence of two compensating diagrams.  
The direct chiral commutator $[J_{05}, \bar q (G\sigma)\gamma_5 q]$ tends to 
cancel against a pion re-scattering diagram (see Ref. \cite{P} for details). 
The resulting expression for $\bar g_{\pi NN}^{(1)}$ is given by 
the following linear combination of nucleon matrix 
elements for $ \bar{q}g_s (G\si) q$ and $\bar qq$:
\be
\fr{\tilde d_u-\tilde d_d}{4f_\pi}\langle N | \uGu+\dGd 
- \tilde m_0^2(\uu+\dd)|N\rangle.
\ee
This vanishes if 
$\langle N | \bar{q}g_s (G\si) q|N\rangle = \tilde m_0^2 \langle N | 
\bar{q} q|N\rangle$, a relation that holds in the vacuum factorization 
approximation. Thus, in order to obtain a non-zero result, one 
has to go beyond the factorization approximation. A 
full QCD sum-rule analysis \cite{P} produces a preferred 
range for this coupling with the ``best'' value given by 
\be
\bar g_{\pi NN}^{(1)} = 2\times 10^{14}~ 
 (\tilde d_u-\tilde d_d)~ {\rm cm}^{-1}.
\ee 
Combined with the relevant nuclear, $S(\gone)$, and atomic, $d_{\rm Hg}(S)$, 
pieces of the calculation, this expression enters the 
first line of Eq.~(\ref{Hgmaster}). 

After this brief review of the color EDM 
contributions to $\bar g_{\pi NN}^{(1)}$, for large $\tan\beta$ we also need to
generalize this derivation to include $\gone(C_{q_1 q_2})$.
Both diagrams, the direct chiral commutator and 
the pion re-scattering diagram, are present in this case as well.
Luckily, in contrast with the case of CEDM sources, vacuum 
factorization does now produce a non-zero result. Combining these two 
diagrams, we find a contribution that corresponds to the 
{\em direct factorization} of the pseudoscalar bilinear $\bar q i \gamma_5 q$
via the chiral commutator with the pion. The final expression for 
$\gone$ takes the form,
\begin{eqnarray}
\bar g_{\pi NN}^{(1)} &=& \fr{\qq}{2f_\pi}\langle N|
C_{dd}\bar dd + C_{sd}\bar ss+ C_{bd}\bar bb |N\rangle \nonumber\\
 &=& -8\times 10^{-2}{\rm GeV}^2\left(\fr{0.5C_{dd}}{m_d} 
+ 3.3\kappa\fr{C_{sd}}{m_s}+ 
\fr{C_{bd}}{m_b}(1-0.25\kappa)\right).
\label{gone}
\end{eqnarray}
Note that within the factorization approximation $C_{db}$, $C_{ds}$, $C_{ss}$ 
and $C_{bb}$ do not contribute to $\gone$. The final expression 
(\ref{gone}) depends on the same matrix element of $\bar ss$ 
over the nucleon as previously encountered in the calculation 
of $C_S$, and parametrized by $\kappa$.

\end{document}